\documentclass[usenatbib]{basi}
\usepackage[T1]{fontenc}
\usepackage[british]{babel}
\usepackage[varg]{txfonts}
\begin{document}
\title[Overview of Gamma-ray Bursts in the Fermi era]{An overview of the
       current understanding of Gamma-ray Bursts in the Fermi era}
\author[P.~N.~Bhat and S.~Guiriec]{P.~N.~Bhat$^1$\thanks{email:
       \texttt{Narayana.Bhat@nasa.gov}} and S.~Guiriec$^{1,2}$\thanks{email:
       \texttt{sylvain.guiriec@nasa.gov}, NASA Postdoctoral Program Fellow}\\
       $^1$University of Alabama in Huntsville, 320 Sparkman Dr.,
       Huntsville, AL, 35805, USA \\
       $^2$NASA, Goddard Space Flight Center, Greenbelt, MD 20771, USA }

\pubyear{2011}
\volume{39}
\pagerange{\pageref{firstpage}--\pageref{lastpage}}

\date{Received 2011 November 07; accepted 2011 November 12}

\maketitle
\label{firstpage}

\begin{abstract}
Gamma-ray bursts are the most luminous explosions in the Universe, and their
origin as well as mechanism are the focus of intense research and debate. More
than three decades since their serendipitous discovery, followed by several
breakthroughs from space-borne and ground-based observations, they remain one of the
most interesting astrophysical phenomena yet to be completely understood. Since
the launch of Fermi with its unprecedented energy band width spanning seven
decades, the study of gamma-ray burst research has entered a new phase. Here we
review the current theoretical understanding and observational highlights of
gamma-ray burst astronomy and point out some of the potential promises of
multi-wavelength observations in view of the upcoming ground based
observational facilities.
\end{abstract}

\section{Introduction}\label{s:intro}

Gamma-ray bursts (GRBs) are short, intense and distant flashes of $\gamma$-rays
that occur at random locations in the sky with their peak power in the 200--500
keV range. During their appearance, they often outshine all other sources combined 
in the $\gamma$-ray sky. Early observations have detected what is now
referred to as the prompt emission which is brief (milliseconds to minutes),
highly variable (in time scales of sub-ms to tens of seconds), non-thermal, and
observed mostly in the keV/MeV energy range. A breakthrough in the field of GRB
research happened with the numerous GRB detections from the Burst And Transient
Source Experiment (BATSE), which flew, with other instruments, on board the
Compton Gamma-Ray Observatory (CGRO). Before BATSE, the distance scale of GRBs
was unknown. The scientific opinion at the time was divided among the various
theories predicting distance scales ranging from our own solar system to the
edges of the known Universe.

\begin{figure}
\centerline{\includegraphics[width=11.0cm]{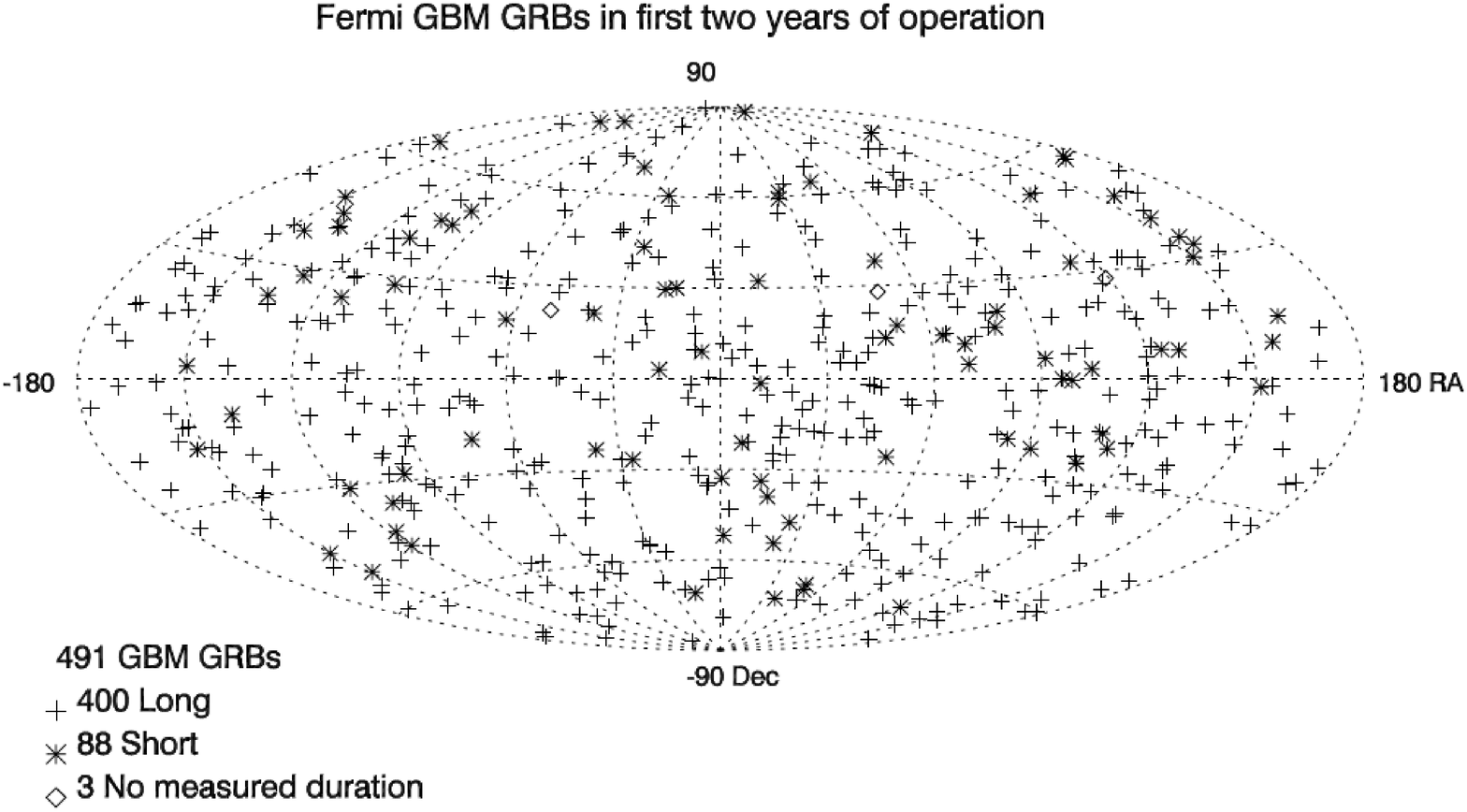}}
\caption{During the first 2 years of operation GBM detected and located 491
GRBs. The locations for 400 long bursts and 88 short bursts are plotted in this
figure separately to demonstrate that the both long and short GRBs are
uniformly distributed in the sky and is consistent with the earlier results of
BATSE experiment underlining their cosmological origin.\label{f:sky_dist}}
\end{figure}

BATSE's improved spatial sensitivity proved that GRBs are isotropically
distributed on the celestial sphere ruling out possible correlations with the
local distribution of stellar or gaseous mass (e.g. our Galaxy, the LMC, M31,
globular clusters, the Virgo cluster) which is not isotropic.
Fig.~\ref{f:sky_dist} shows a sky distribution of GRBs from the recent 
Gamma-ray Burst Monitor (GBM) experiment (see \S \ref{ss:Fermi_Instr} for details).

This piece of evidence strongly suggested a cosmological origin for GRBs
although an extended halo around our galaxy could still generate a uniform
distribution similar to the one observed. Key GRB properties have emerged as
soon as their cosmological nature has been established by BATSE
\citep{1995ARAA...33..415}. GRBs have been thought as probes of the processes
and environments of star formation out to the earliest cosmic epochs.

\begin{table}
\caption{List of past and operating space borne instruments that have detected
GRBs along with the approximate numbers of GRBs detected by them as of August
2011.}\label{tab_missions}
\medskip
\centering
\def\hpad{\kern15pt}
\begin{tabular}{cccr}\hline
Instrument & Observing  energy range & Period of operation & GRBs detected\\\hline

Vela 3-6   & 200 keV -- 1 MeV             & 1967 -- 1979  &  100 \hpad \\[4pt]
CGRO       & 50 -- 300 keV (BATSE)        & 1991 -- 2000  & 2704 \hpad \\[-2pt]
           & 20 MeV -- 30 GeV (EGRET)     &               &    5 \hpad \\[4pt]
Konus/WIND & 10 keV -- 10 MeV             & 1994 -- now   & 2114 \hpad \\[4pt]
BeppoSAX   & 2 -- 600 keV                 & 1996 -- 2002  & 1082 \hpad \\[4pt]
HETE-2     & 2 -- 400 keV                 & 2000 -- 2006  &  104 \hpad \\[4pt]
Integral   & 15 keV -- 10 MeV             & 2002 -- now   &  760 \hpad \\[4pt]
Swift      & 15 -- 150 keV (BAT)          & 2005 -- now   &  612 \hpad \\[-2pt]
           & 0.3 -- 10 keV (XRT)          &               &            \\[-2pt]
           & 170 -- 600 nm (UVOT)         &               &            \\[4pt]
AGILE      & 10 keV -- 700 keV (SA+MCAL)  & 2007 -- now   &  210 \hpad \\[-2pt]
           & 30 MeV -- 30 GeV (GRID)      &               &    3 \hpad \\[4pt]
Fermi      & 10 keV -- 40 MeV (GBM)       & 2008 -- now   &  765 \hpad \\[-2pt]
           & 20 MeV -- 300 GeV (LAT)      &               &   28 \hpad \\\hline
\end{tabular}
\end{table}

After CGRO the Russian Konus experiment \citep{1995SSR...71..265} on board the
Wind satellite launched in November 1994 as well as High Energy Transient
Explorer (HETE-2) experiment \citep{2003AIP...662..3} launched in October 2000
recorded more than 2000 GRBs \citep{2003AIP...662..143, 2004AIP...727..57}
adding to the earlier record of more than 2700 GRBs recorded by BATSE. HETE-2
made the first observations of long GRBs spatially associated with Type Ic
supernovae. More recently, the Swift satellite is a currently operating mission
with its Burst Alert Telescope (BAT) that detects and locates GRBs within
$0.1^\circ$. In addition, it also carries an X-ray Telescope (XRT) and an
Ultra-Violet and Optical Telescope (UVOT) which when slewed to the burst location
can successfully detect the GRB afterglows and locate the source with an
accuracy of arcseconds. The success of the Swift mission has resulted in the
measurement of the redshifts of nearly 200 GRB hosts out of a total of 600 GRBs
detected.
%
%
Swift made major breakthroughs in the understanding of the GRB afterglow
emission that were hitherto unknown. Table~\ref{tab_missions} summarises the
satellite observations and the on board $\gamma$-ray/X-ray detector
characteristics as well as the total number of bursts detected by them as of
September 2011 \citep[updated version of the table in][]{2010Astro...1012..0558}.

The precise location of GRBs is a challenge since GRBs are relatively short
transient events which do not repeat and that they occur randomly in time and
space. As a result, a majority of the early GRB missions before BATSE could not
localise GRBs. BATSE was the first mission which could detect and locate a GRB
within a couple of degrees. The angular resolution of BATSE was still too
coarse to repoint optical or X-ray telescopes to the location of the explosion
(which have a small field-of-view necessitating a few arc-minute location
accuracy) to search for a burst counterpart the possible existence of which was
theoretically predicted at the time \citep{1992MNRAS...258..41,
1993APJ...405..278}. Early attempt to locate a GRB precisely was made by the
Inter-Planetary Network
(IPN\footnote{\texttt{http://www.ssl.berkeley.edu/ipn3/}}) which consists of a
group of spacecrafts (so far 27 spacecrafts have participated in this network)
equipped with $\gamma$-ray detectors. By timing the arrival of $\gamma$-rays
from a burst at several spacecrafts the burst could be localised with an
accuracy depending on the number of satellites detecting it. However very few
bursts were located by this technique and even when successful, the time delay
was too long to be useful for follow-up observations at other wavelengths.

A breakthrough happened in early 1997, with the Dutch/Italian satellite
BeppoSAX which was equipped with a wide angle GRB detector and X-ray telescopes
with coded aperture. As soon as a GRB was detected by the on board wide-field
camera (WFC) the satellite could slew such that the GRB source is in the field
of view of the X-ray detectors which could then locate it, within hours, with
in a few arcmin accuracy. This allowed the detection of the first fading X-ray
emission from a long burst GRB\,970228 leading to the improved localisation of
the source accurate enough to facilitate follow-up observations at optical
wavelengths for the first time. These observations initially identified a
fading optical counterpart \citep{1997Nat...386..686}, and, after the burst had
faded, long duration deep imaging identified a distant host galaxy with a
redshift $z = 0.498$ at the location of the burst. The redshifts are usually
measured from the emission lines or the absorption features of the host
galaxies imposed on the afterglow continuum. GRB\,970508 was the first GRB for
which the afterglow was concurrently observed over the whole electromagnetic
spectrum a few hours after the burst. This was also the first GRB whose
distance was estimated by measuring its spectroscopic redshift
\citep{1997Nat...387..878}. BeppoSAX thus paved the way for detections of more
host galaxies and redshifts through spectroscopy of the GRB host galaxies, and
settled the question of cosmological origin for long GRBs once and for all. The
distance measurement of GRBs opened the door to the study of intrinsic
properties of their sources for the first time. The typical $\gamma$-ray
fluences measured by BATSE are of the order of $10^{-5}$ erg~cm$^{-2}$ which
translates to an equivalent isotropically-emitted energy of $E_{\rm iso} \sim
10^{53}$ erg at the source. The total $\gamma$-ray fluence can be significantly
smaller if the emission is beamed which is most likely the case.

\begin{figure}
\centerline{\includegraphics[width=6.5cm]{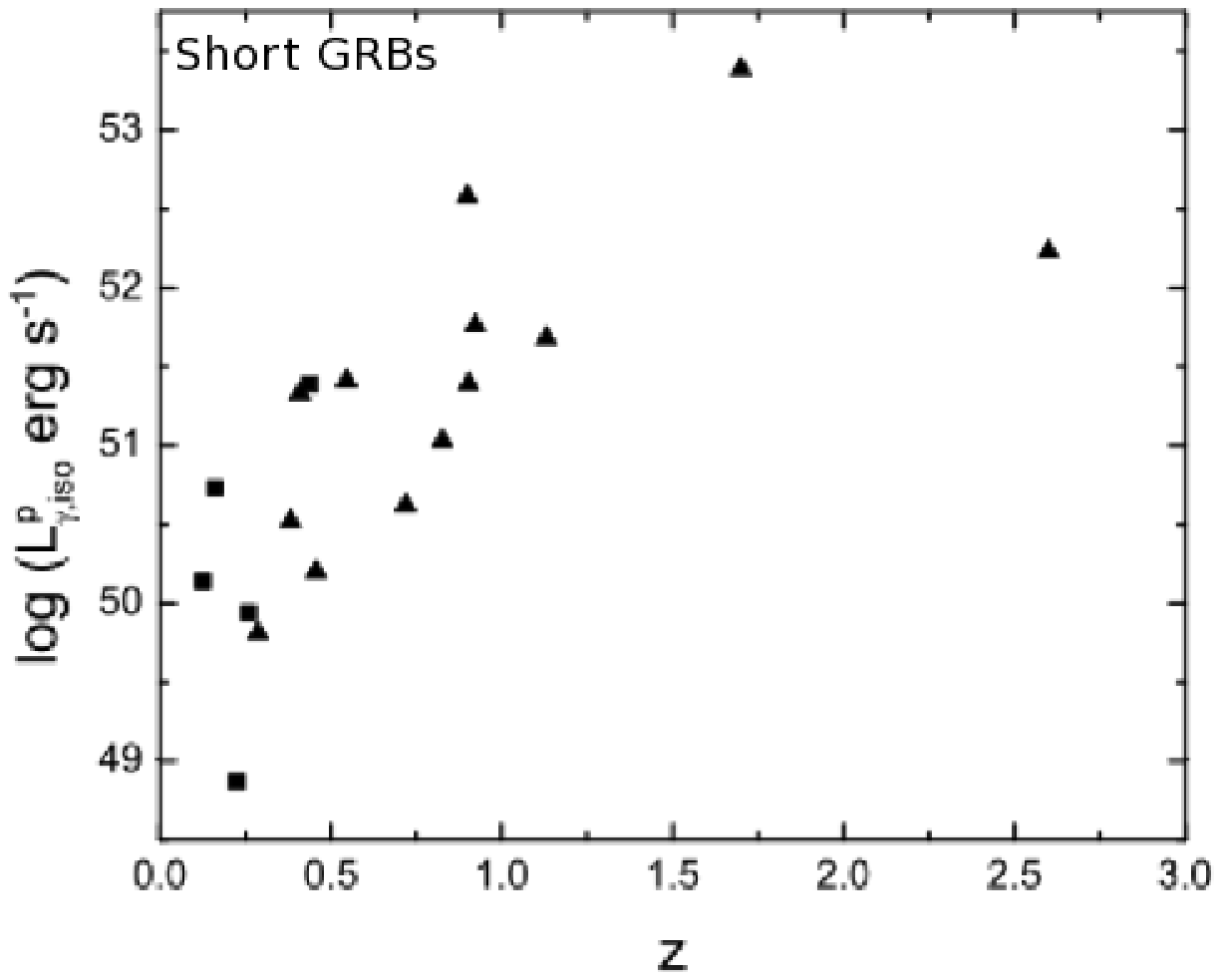}
            \includegraphics[width=6.5cm]{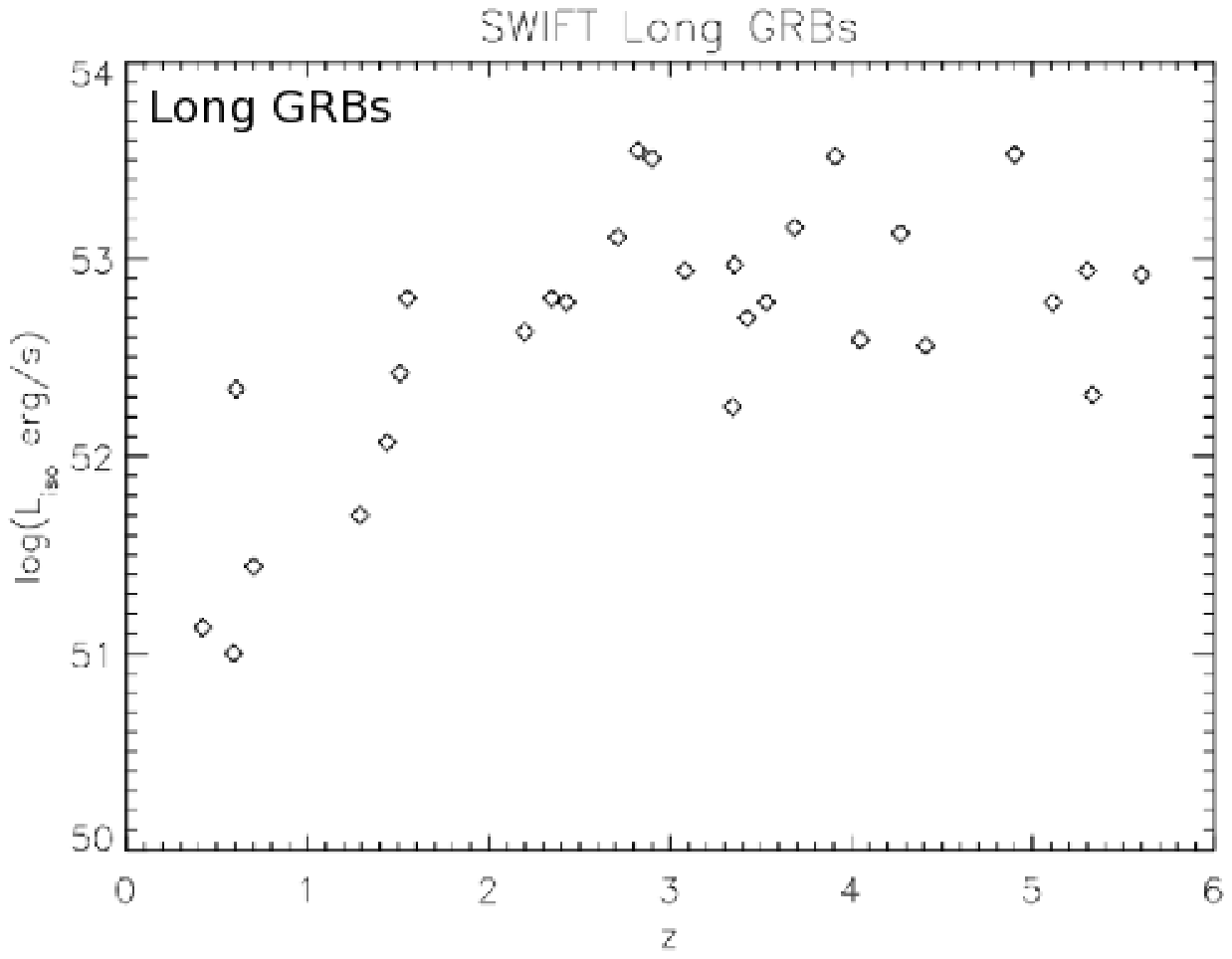}}
\caption{Two-dimensional luminosity-redshift distributions of short
\citep{2009MNRAS.392...91} and long \citep{2007MNRAS.382..342} GRBs using the
recent Swift observations. The very high redshift GRB\,090423 at $z \sim 8.2$
\citep{2009Nat...461.1254} or GRB\,090429B at $z \sim 9.4$
\citep{2011ApJ...736....7} has not been included in this plot.\label{f:z_dist}}
\end{figure}

Fig.~\ref{f:z_dist} shows the isotropic energy distribution of GRBs as a
function of redshift for short and long GRBs. Considering the limited number of
short GRBs with redshifts the two distributions do not seem to be very
dissimilar (see \S\ref{ss:t90} for details on burst classification).

In \S\ref{s:theory} the current theoretical understanding of GRB prompt and
afterglow emissions are briefly explained. While \S\ref{ss:cgro} gives a brief
description of the observational status prior to the launch of Fermi, \S
\ref{s:Fermi} describes the Fermi instrumentation and its synergy with other
$\gamma$-ray missions currently operating. \S\ref{ss:temp_res} and
\ref{ss:spec_ana} describe the latest observational results relating to both
temporal and spectroscopic aspects of GRBs respectively. In \S\ref{s:inf_res}
a brief description is given of the inferred results on GRBs as a class as well
as some applications of Fermi GRB results relating to the attenuation of high
energy $\gamma$-rays from distant sources and fundamental physics. Finally the
paper concludes by identifying some of the important prospects of future
breakthroughs in the understanding of GRB astrophysics when the new ground
based multi-wavelength observatories become operational. In the next section we
plan to present the basic understanding of what is generally believed to be
responsible for the GRB phenomenon primarily for the benefit of beginners in
the field.

\section{Theoretical understanding of GRB physics}\label{s:theory}

\subsection{Progenitor and central engine}\label{ss:fireball}

In the following sections we outline the observations that support these
theoretical conclusions as well as those that often challenge them. Our
understanding of the physics of GRBs is still far from definitive; however,
there is a general agreement in the community that the progenitors of the long
duration ($T_{90}>2$~s) bursts are associated with the deaths of massive stars
in a specific kind of supernova-like event commonly referred to as a collapsar
or hypernova. A merger of two compact objects, such as two neutron stars or a
neutron star and a stellar mass black hole is believed to be giving rise to
short duration ($T_{90} \leq 2$~s) GRBs. For a more detailed discussion on
burst durations see Section \ref{ss:t90}.

\subsection{Fireball model}

Several models have been proposed to explain the $\gamma$-ray emission
mechanisms following the cataclysmic events mentioned above. Observational data
outlined so far led over the years to the development of the well accepted GRB
`standard model' that describes the main properties of the GRBs with standard
physics applied to somewhat `exotic objects'. The widely accepted
interpretation of the GRB phenomenology is that the observable effects are due
to the dissipation of the kinetic energy of a relativistic expanding wind, a
`fireball', regardless of the nature of the underlying central engine
\citep{1978MNRAS...183..359}. No known process in the Universe can produce this
much energy in such a short time. The GRB central engine, resulting from the
collapse, produces collimated jet mainly composed of electrons, positrons,
photons and a small amount of baryons moving at relativistic speeds. The
outflow wind is believed to be highly inhomogeneous and the high density
regions can be assimilated to separate shells propagating at various
velocities. The observed prompt emission in the $\gamma$-ray energy band would
be mostly synchrotron radiation produced by the radially propagating electrons
accelerated during collisionless shocks inside these jets.

\subsection{Relativistic expansion and pair opacity}\label{ss:relex}

In this section we describe why the observation of GRB prompt emission can only
be explained by highly relativistic outflows. Rapid rise times as short as
0.2~ms \citep{1992NAT...359..217} and durations as short as 1~ms
\citep{1994APJS...92..229} imply that the observed emission regions are
compact, $r_0 \sim 10^7$~cm. In addition, the high $\gamma$-ray luminosities
observed by the detectors coupled to the cosmological distances of the source,
$L_\gamma \sim 10^{52}$ erg~s$^{-1}$, should result in a very high opacity of
the GRB wind to $\gamma$-rays due to pair creation since the energy of observed
$\gamma$-ray photons is above the threshold for pair-production. The density of
photons at the source $n_\gamma$, is approximately given by
\[
   L_\gamma = 4\pi {r_0}^2 c n_\gamma \epsilon,
\]
where $\epsilon \simeq 1$~MeV is the characteristic photon energy. Using $r_0
\sim 10^7$\,cm, the optical depth for pair production at the source is
\[
  \tau_{\gamma\gamma} \sim r_0n_\gamma\sigma_{\rm T}
    \sim \frac{\sigma_{\rm T} L_\gamma}{4\pi r_0 c \epsilon} \sim 10^{15}
\]
where $\sigma_{\rm T}$ the Thompson cross-section. This is obviously in
contradiction with the observation since a large amount of photons with energy
above the threshold are actually detected by telescopes. This is the so-called
`compactness problem' in $\gamma$-ray burst astrophysics. The observation of
$\gamma$-ray flux above $\sim 1$~MeV in several GRBs shows that pair creation
is not predominant in the emission wind implying a low density of photons above
the pair creation threshold at the source. This problem can be solved when we
consider an expanding relativistic wind. Due to relativistic effect, photons
below the pair creation threshold in the comoving frame are blue shifted and
can appear to be above the pair-production threshold in the observer frame.
Relativistic expansion thus provides a very efficient way of reducing the rate
of pair creation in the moving source frame since the photons are softer by a
factor of $\Gamma$, where $\Gamma$ is the Lorentz factor of the relativistic
flow. In addition the size of the emission region becomes $\Gamma^2 c\Delta t$
and the density of photons is thus reduced considerably ($\Delta t$ is the
observed variability time scale). Moreover, relativistic beaming implies that
we observe only a small fraction $1/\Gamma$ of the source irrespective of the
opening angle of the jet. Hence the angle at which the photons collide must be
less than the inverse of the bulk Lorentz factor $\Gamma^{-1}$ of the
relativistic flow which drastically reduces the effective pair production
cross-section if $\Gamma$ is large. As a result, for a differential photon
spectrum, $E^{-\alpha}$, the source becomes optically thin if $\Gamma \geq 100$
assuming $\alpha \sim 2$ \citep{1999PR...314..575, 2001APJ...555..540}. A
general expression for the lower limit of the bulk Lorentz factor so that the
source is transparent to a $\gamma$-ray photon of observed energy $E_\gamma$ is
\citep{2003LNP...598..393}.
\[
  \Gamma \geq 250 \left[
    \left(\frac{L_\gamma}{10^{52}\:{\rm erg~s}^{-1}}\right)
    \left(\frac{E_\gamma}{100~{\rm MeV}}\right)
    \left(\frac{\Delta t}{10^{-2}\:{\rm s}}\right)
  \right]^\frac{1}{6}
\]
 More recenly, refined estimates of the opacity to pair production were proposed considering time-dependent effects  in a simplified single zone case  \citep{2008ApJ...677...92} and in a more realistic multi-zone case that leads to transparency to high energy $\gamma$-rays at a significantly lower values of $\Gamma$  \citep{arXiv:1107.5737} .
However, the most energetic GRB photons whose energies are above $E_\gamma$ may
still suffer from absorption leading to pair production in the GRB winds
\citep{2003APJ...599..380}. This may lead to energy dependent optical depth
consequently giving rise to the delayed emission of higher energy $\gamma$-rays
\citep{2005ApJ...633..1018}. Further discussion on this is in
\S\ref{sss:he_delay}.

\subsection{Fireball evolution}\label{ss:fb_evol}

A fireball is essentially a dynamic object whose properties quickly evolve with
time. It can be characterised by an initial energy $E_0$ and a radius $R_{\rm
in}$, while the energy to mass ratio, $\eta = E_0/M_0c^2$ with $M_0 \ll
E_0/c^2$, $M_0$ represents the baryon loading factor of the fireball and $\eta$
the mean energy per baryon.

\subsubsection{Acceleration}

The initial optical depth being extremely high as mentioned before, the radial
expansion of the fireball is a consequence of the highly super-Eddington
luminosity where the internal energy is converted into kinetic energy. As the
fireball cools during its expansion, its temperature varies as $T_\gamma
\propto R^{-1}$. Since the total energy $E_0$ is constant, the bulk Lorentz
factor increases linearly with $R$, $\Gamma \propto R$, until it saturates at a
value $\Gamma_{\rm max} = \eta R_{\rm in}$. The radius of the fireball, where
the bulk Lorentz factor reaches $\Gamma_{\rm max}$, is called the saturation
radius, $R_{\rm s}$. Beyond the saturation radius, the bulk Lorentz factor
$\Gamma$ coasts at this value of $\Gamma_{\rm max}$. In this initial phase, the
wind is optically thick and it is usually believed to become optically thin at
the photospheric radius ($R_{\rm ph} \approx 10^{11}-10^{12}$\,cm). Before
$R_{\rm ph}$, the photons and the electrons are in thermal equilibrium and the
electron velocity distribution is Maxwellian. Beyond the saturation radius, the
shells propagate in the jet at the constant velocities until the broadening
radius ($R_{\rm is}$) where the shells expand indicating the beginning of the
so-called internal shock phase \citep{1994ApJ...430..L93, 1992ApJ...395..L83,
1994ApJ...427..708}.

\subsubsection{Internal shocks}

During the internal shock phase ($R_{\rm is}\approx 10^{14}-10^{15}$\,cm),
faster shells catch up with slower ones leading to mildly relativistic
collisionless diffusive shocks where the charged particles present in the shock
region are accelerated. During the internal shock phase, charged particles with
enough energy would be accelerated by scattering off the magnetic perturbations
in the up stream or down stream shells. This acceleration mechanism is called
the second order Fermi process. During this process the charged particles may
cross the shock front multiple times where they are accelerated through first
order Fermi processes. The first order Fermi process is much more efficient
than the second order and the former process imparts the same amount of energy
to the charged particle whether it crossed the front upstream or downstream. A
charged particle performs a Fermi cycle when crossing the shock front twice. It
can take several Fermi cycles before gaining too much energy and escape.

If the photospheric radius $R_{\rm ph} < R_{\rm is}$, then the wind becomes
optically thin before the internal shock phase starts, and a thermal emission
can be observed first \citep{1994ApJ...430..L93}. During the internal shock
phase, the charged particles (mostly electrons) are accelerated through Fermi
processes as discussed above. The electron energy distribution resulting from
Fermi acceleration is a power law, and the final electron distribution would be
a Maxwellian deformed above the Fermi energy threshold and with an extended
power law tail. The $\gamma$-ray emission observed during the prompt phase of
GRBs would be mostly the synchrotron emission from the electrons which
propagate and are accelerated within an intense magnetic field. Each peak seen
in the prompt emission light curves would correspond to an individual shock
resulting from the collision of a pair of shells with unequal Lorentz factors.
Protons could also be accelerated in these shocks, however, the acceleration
time is much longer than that for electrons. Protons thus accelerated could
produce synchrotron emission at higher energies. If $R_{\rm ph} > R_{\rm is}$,
then the internal shock phase starts while the wind is still optically thick.
This would result in heating the shells which in turn results in changing the
electron distribution and the temperature of the photosphere while the jet
becomes optically thin.

Thus, the variability of the light curve reflects the variable activity of the
central engine. The total duration of the GRB corresponds to the total duration
of the central engine activity. The main advantage of this model is that
internal shocks are expected naturally in the baryonic outflow and they can
easily explain the rapidly variable and diverse light curves of prompt
$\gamma$-ray emission. In addition, the simplest interpretation of the emission
as synchrotron radiation is roughly in agreement with the observed burst prompt
emission spectra. The electrons have typical Lorentz factors in the range
$10^2{-}10^3$ and gyrate in a $\sim 10^6$~G magnetic field producing $\sim
100$~keV synchrotron emission \citep{2007PR...442..166}. The main disadvantage
of internal shock model is its efficiency. Internal shocks are expected to
radiate only a small fraction ($\sim 1\%$) of total energy flow
\citep{1998MNRAS...296..275} whereas the observations suggest that the
$\gamma$-ray efficiencies are much higher \citep{2006ApJ...638..354}.

During the internal shock phase, the width of the shells expands as the radius
increases reducing the density of the magnetic perturbations they contain.
Then, shocks occurring far from the central engine are less efficient leading
to particles with lower energy. In addition, the magnetic field strength
decreases with the distance to the central engine, reducing the intensity and
the maximum energy of the synchrotron emission.

\subsubsection{External shock}

The internal shock phase ends at the decelerating radius $R_{\rm
es}$ ($10^{16}{-}10^{17}$~cm), when the wind is slowed down by interacting with
the interstellar medium in a relativistic shock, called external shock. The
external shock would be composed of two shocks, the forward shock which expands
in the direction of the interstellar medium, and the reverse shock which comes
back in direction of the central engine. The position of $R_{\rm es}$ depends
on the density of the interstellar medium. In the case of the compact object
merger, GRBs are believed to happen in low interstellar density environment
(compact objects can migrate far from the star forming region before merging),
while in the hypernova scenario GRBs occur mostly in the star forming region
where the interstellar density is high. The surroundings of a hypernova are
enriched by the stellar wind of the star before the collapse. $R_{\rm es}$ is
therefore believed to be larger for the merger scenario than that for the
hypernova. During the external shock, charged particles of the interstellar
medium and of the wind would be accelerated, and would radiate through
synchrotron and/or inverse-Compton scattering process, which would be
responsible for the afterglow emission observed from radio wavelengths to
X-rays and maybe $\gamma$-rays (see \S\ref{ss:afterglow} for more details).

External shocks could also produce the observed prompt emission according to
some models. In these models radiation pulses are emitted when a relativistic
shell ejected by the GRB central engine is decelerated by the circum-burst
material \citep{1993APJ...405..278}. A homogeneous medium leads to a single
pulse but an irregular, clumpy environment can produce a complex profile if a
large number of small clouds are present \citep{1999ApJ...513..L5}. One major
difficulty with this model is to explain the rapid variability observed in GRB
light curves at various energies. In this case the clumpiness has to be of the
order of $\sim \Gamma_{\rm f} c \Delta t/(1+z)$ where $\Delta t$ is the
variability time scale observed in the GRB light curve and $\Gamma_{\rm f}$ is
the Lorentz factor of the forward shock. However this process has been shown to
be very inefficient \citep{1997ApJ...485..270}. Variability can be recovered
while maintaining high efficiency if the shell that moves with a bulk Lorentz
factor $\Gamma_{\rm f}$ contains emitting clumps.

The external shock model has been successfully used for explaining the GRB
afterglow emission observed from hours to days from radio wavelengths to X-rays
and $\gamma$-rays following the prompt emission, and even for years at radio
wavelengths. The GRB afterglow emission is presented in detail in \S
\ref{ss:afterglow}.

\begin{figure}
\centerline{\includegraphics[width=11.0cm]{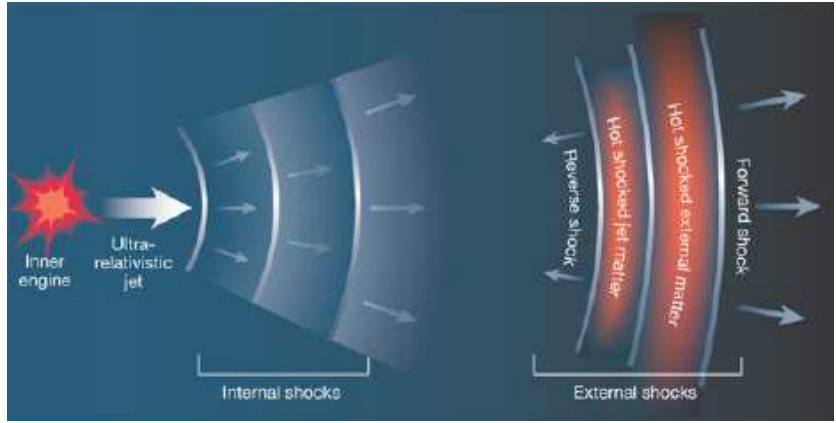}}
\caption{Sketch showing the different phases involved in the `fireball' model
with internal shocks producing the $\gamma$-ray prompt emission and external
shock with the interstellar medium or the star wind responsible for the
afterglow phase observed in radio, optical, X-rays, $\gamma$-rays
\citep{2003Nat...422..268}.\label{f:grb}}
\end{figure}

Fig.~\ref{f:grb} shows a sketch of the $\gamma$-ray burst production
mechanism, the prompt emission regions during the internal shocks  as well as
the afterglow emission from the external shocks.

\subsection{Prompt emission from a magnetised flow}\label{ss:mag_flow}

There are several alternative models for the prompt GRB emission, which so far
have not found wide use for explaining the observations. The most plausible of
these, despite the technical difficulties which impair its applicability,
considers the main $\gamma$-ray burst emission to arise from magnetic
reconnection or dissipation processes, if the ejecta is highly magnetised or
Poynting flux dominated. If the outflow energy is carried mostly by Poynting
flux until where the interaction with the external medium starts, the
dissipation of the bulk energy into internal energy cannot occur through
internal shocks. In this case the energy source of the radiation is most likely
magnetic dissipation. \citet{2003MNRAS...346..540} suggest that electromagnetic
current-driven instabilities dissipate magnetic energy into heat and high
energy particles. The propagation of these high energy particles within a
strong magnetic field, is the source of the observed prompt $\gamma$-ray
emission.

\citet{2006ApJ...651..333} suggests a model which is a flow consisting of a
combination of a strong magnetic field and pair-radiation plasma. In this model
the radius in which the outflow becomes optically thin is determined by the
pair enrichment of the ambient medium through the interaction with the flow
radiation. The prompt emission is generated by the inverse Compton scattering
of the seed photons that are carried in the flow by pairs which are accelerated
by the reconnecting magnetic field. The energy at the peak of the prompt
emission spectrum reflects the temperature of these seed photons Lorentz
boosted to the observer frame.

\subsection{GRB afterglow  emission}\label{ss:afterglow}

The prompt emission is usually followed by a longer-lived `afterglow' emitted
at longer wave\-lengths. Afterglows are now known to be broad-band, having been
detected in the X-ray, the optical/infrared and the radio bands. X-ray
afterglows are mostly decaying when they are detected. Optical afterglows are
generally decaying, with an initial early rising lightcurve having been caught
in only a few bursts (e.g.\ GRB\,970508, GRB\,990123). In each band, the
general power-law decay behaviour can be summarised as: $t^\alpha \nu^\beta$
where $\alpha\sim -0.9 (-1.0)$ and $\beta \sim -1.4 (0.7)$ for X-ray
\citep{2001SBH} (optical, \citealt{2006ApJ...637..889}) afterglows. GRB\,970228 was the
first GRB for which an X-ray afterglow was observed \citep{1997Nat...387..783}.
The X-ray emission was fading fast following a power law $F(t) \sim
T^{-1.3 \pm 0.1}$. This also led to the detection of the first optical afterglow
from this GRB \citep{1997Nat...386..686}. Very often there are various types of
deviations from the simple power law decay. These include steepenings, bumps
and wiggles (e.g.\ GRB\,021004; GRB\,030329); Essentially every GRB with an
afterglow detection has an underlying host galaxy. The GRB host galaxy
properties (e.g. magnitude, redshift distribution, morphologies) are typical of
normal, faint, star forming galaxies \citep{2003SPIE.4834..238}. The GRB
afterglow's positional offsets with respect to the host galaxy are consistent
with GRBs being associated with the star forming regions in the galaxies
\citep{1998ApJ...507..L25}. In the radio band, on the other hand, the spectral
index is generally positive for the observations which are typically in the 5
and 8.5 GHz bands. The light curves usually do not follow a simple power law
decline \citep{2003AJ...125..2299}. Some sources can be observed on timescales
of years, and a late-time flattening (with respect to the standard fireball
model) is often observed \citep{2003Aph...0308189}.

\subsubsection{Standard afterglow model}\label{sss:std_ag_model}

The standard afterglow model assumes a highly relativistic expansion of a
spherical outflow in the adiabatic regime into a homogeneous external medium.
The external shocks appear when the relativistic outflow is slowed down in the
interstellar medium (ISM) surrounding the source or the stratified wind ejected
by the progenitor star prior to the collapse \citep{1992MNRAS...258..41}.
Generally two shocks form: an outgoing shock, called the forward shock, that
propagates into the surrounding medium and a reverse shock that propagates back
into the ejecta. The external shock successfully explains multi-wavelength
afterglow radiation which begins at a distance where most of the energy of the
ejecta is transferred to the medium. However the assumptions in the standard
model are too simplistic while the real life situations more complex. The
impact of the reverse shock is invoked to explain the early optical flashes
while the beaming of the outflow within a jet of solid angle $\Omega_{\rm j}$
is an ingredient to reduce the energetics of GRBs as mentioned before.
%

In order to produce the observed spectrum during both the afterglow and the
prompt emission phases, electrons must be accelerated in the collisionless
shocks to a power law distribution, $dn_{\rm e}/d\Gamma_{\rm e} \propto
\Gamma_{\rm e}^{-p}$ with $p \simeq 2$ where $\Gamma_{\rm e}$ is the electron
Lorentz factor. Such a distribution is expected in the internal shocks, which
are mildly relativistic. Numeric and analytic calculations of particle
acceleration via the first order Fermi mechanism in relativistic shocks show
that similar indices, $p \approx 2.2$, are obtained for highly relativistic
shocks as well \citep{1998PRL...80..3911, 2000ApJ...542..235}. The spectrum of
radiation is likely to be due to synchrotron radiation, whose peak frequency in
the observer frame is $\nu_m \propto \Gamma_{\rm max} B^\prime \Gamma_{\rm
e}^2$, where the comoving magnetic field $B^\prime$ and electron Lorentz factor
$\Gamma_{\rm e}$ are likely to be proportional to $\Gamma_{\rm max}$ (maximum
value of the bulk Lorentz factor). This implies that as $\Gamma_{\rm max}$
decreases, so will $\nu_m$, and the radiation will move to longer wavelengths.
Consequently, the burst would leave a radio remnant weeks after the explosion
\citep{1993ApJ...418..L5}. The observation of linear polarisation at the few
percent level observed in a number of optical or IR afterglows
\citep{2000ARAA...38..379} supports the paradigm of synchrotron emission as the
dominant emission mechanism in the afterglow.

\subsubsection{Prompt flashes and reverse shocks}\label{sss:rebright_rs}

The first optical brightening following a $\gamma$-ray burst was observed from
GRB\,970508. More recently the observation of extremely bright ($m_v \sim 9$)
optical flash in the burst GRB\,990123 which is interpreted as due to reverse
shock \citep{1999ApJ...517..L109} even though a prompt optical flash could be
expected from either internal shock or reverse shock of external shock. The
decay rate of the optical flux from the reverse shock is much faster (and that
from internal shock is even faster) than that of the forward shock. Generally
reverse shock is expected to be mildly relativistic and hence radiate much
softer radiation than the forward shock \citep{2006RPPh...69..2259}. After the
launch of Swift, new prompt optical observations with robotic telescopes have
greatly added to the phenomenology of prompt flashes which are consistent with
the reverse shock interpretation.

Currently, Swift is the only mission designed to detect GRB afterglows. The BAT,
with its large field of view of 2 steradians, can detect and locate a GRB, as
well as compute burst positions on board with arc-minute positional accuracy.
The spacecraft then slews to this position and the XRT takes images to obtain
spectra of GRB afterglows. The images are used for higher accuracy position
localisations, while light curves are used to study flaring and long-term decay
of the X-ray afterglow. At the same time the UVOT also takes images and obtains
spectra (via a grism filter) of GRB afterglows. The images are used for
0.5 arcsecond position localisations and following the temporal evolution of
the UV/optical afterglow. Spectra are taken for the brightest UV/optical
afterglows, which can then be used to determine the redshift via the observed
wavelength of the Lyman-alpha cut-off. Accurate localisations are transmitted
in real time to enable ground based observatories to carry out multi-wavelength
follow-up observations of the GRB afterglows.

\begin{figure}
\centerline{\includegraphics[width=11.0cm]{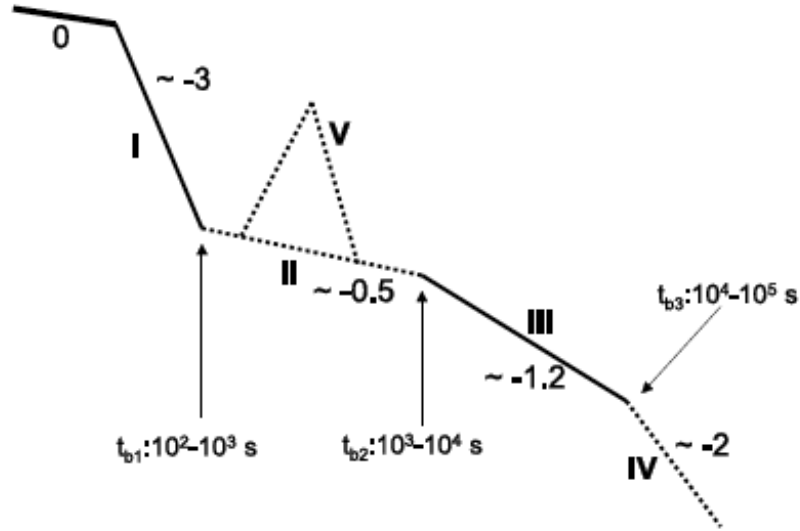}}
\caption{Sketch of an afterglow light curve based on Swift's observations.
Phase `0', corresponds to the end of the prompt emission. Four power-law
light-curve segments together with a flaring component are identified in the
afterglow phase. The components marked with solid lines are the features common
to most long GRBs, while the ones marked with dashed lines are observed in only
a fraction of GRBs. The phase II is often called the plateau. The typical
spectral indices of the power law decay are shown for each segment. The break
between regions III and IV occurs simultaneously for several observed
frequencies (achromatic break) and is related to the geometry of the GRB
relativistic jets. For some bursts the break is not
achromatic.\label{f:swift_xag}}
\end{figure}

Fig.~\ref{f:swift_xag} is a sketch of a typical afterglow observed by Swift
\citep{2006ApJ...642..354}. It consists of a fast decay phase (I) followed by a
plateau region (II), which is a major Swift discovery, lasts for a few thousand
seconds and then reaches typical afterglow decay with a slope of $\sim -1.2$
(III) already observed earlier.

\subsection{Measurement of the jet opening angle}

The spherical assumption is valid even when considering a relativistic outflow
collimated within some jet of solid angle $\Omega_{\rm j} < 4\pi$, provided the
observer line of sight is inside this angle, and $\Gamma_{\rm e} \gtrsim
\Omega^{-1/2}$ \citep{1993ApJ...415..181} so the light-cone is inside the jet
boundary and the observer is unaware of what is outside the jet. However, as
the ejecta is decelerated, the Lorentz factor eventually drops below this
value, and a change is expected in the light curves \citep{1999ApJ...525..737}.
It is thought that this is what gives rise to the achromatic breaks seen in the
light curves of many optical afterglows \citep{2001ApJ...562..L55}

The jet opening angle can be obtained from the observer time $t_{\rm j}$ at
which the flux $F_\nu$ decay rate achromatically changes to a steeper value,
assuming that this corresponds to a causal angle $\Gamma_{\rm e}(t)^{-1}$
having become comparable to the jet half-angle and larger later on. However the
detailed simulation studies show that for off-axis observers, the observable
jet break can be delayed up to several weeks, potentially leading to
overestimation of the beaming-corrected total energy
\citep{2010ApJ...722..235}. In addition, achromatic jet breaks are relatively
rare in the literature. Different explanations for the lack of truly achromatic
breaks have been put forth. It was shown recently that in general $\gamma$-ray
burst afterglow jet breaks are chromatic across the self-absorption break
\citep{2011MNRAS...410..2016}.

\subsection{GRB as UHECR sources}\label{ss:uhecr}

One of the outstanding problems in astronomy is the origin of ultra-high energy
cosmic rays (UHECR). The local ($z=0$) energy production rate in $\gamma$-rays
by GRBs is roughly given by the product of the characteristic GRB $\gamma$-ray
energy, $E \approx 10^{53}$ erg and the local GRB rate. Under the assumption
that the GRB rate evolution is similar to the star formation rate evolution,
the local GRB rate is $\sim 0.5$ Gpc$^{-3}$yr$^{-1}$, implying a local
$\gamma$-ray energy generation rate $\approx~10^{44}$ erg~Mpc$^{-3}$~yr$^{-1}$.
The energy observed in $\gamma$-rays reflects the fireball energy in
accelerated electrons. Thus, if accelerated electrons and protons carry similar
energy (as indicated by afterglow observations) then the GRB production rate of
high energy protons is remarkably similar to that required to account for the
flux of $>~10^{19}$ eV cosmic rays \citep{2003LNP...598..393}. Based on these
arguments GRBs have been proposed as a likely source of UHECRs
\citep{1995PRL...75..386, 2002ApJ...574..65}. However there are some questions
unanswered. These include the limitations to the highest energy attainable by
protons around the bursts' shocks, the spectral slope at the highest energies,
the total energy released in non-thermal particles, and the occurrence of
doublets and a triplet in the data reported by the Akeno Giant Air Shower Array
(AGASA). Considering the uncertainties and the apparent agreement in the energy
budget GRBs seem to be a strong candidate for the source of UHECRs. In
addition, the computed resulting particle spectrum at Earth, fits the 
High Resolution Fly's Eye (HiRes) and AGASA data to within statistical 
uncertainties \citep{2003ApJ...592..378}.

\section{Observational status: pre-Fermi era}\label{ss:cgro}

\subsection{Prompt emission durations and two GRB classes}\label{ss:t90}

\begin{figure}
\centerline{\includegraphics[width=11.0cm]{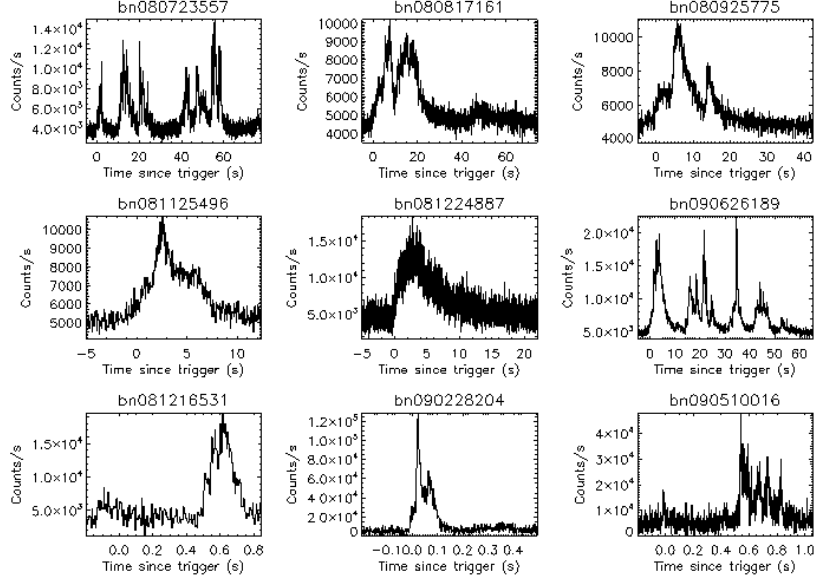}}
\caption{A sample GRB light curves from the GBM experiment demonstrating their
diversity and uniqueness. The first two rows of light curves are from long GRBs
($T_{90}~>~2$~s) while the last row shows light curves from short GRBs ($T_{90}
\leq 2$~s).\label{f:grb_lcs}}
\end{figure}

GRB time profile is unique and unpredictable. Some are smooth and exhibit a
fast rise and exponential decay (often referred to as `FRED') type profile
while others are highly variable with several overlapping narrow pulses (see
Fig.~\ref{f:grb_lcs}). GRBs exhibiting FRED type light curves are shown to be
not any different from other type of GRBs \citep{1994ApJ...426..604}. One clear
example of uniqueness was demonstrated by the lack of success in a search for
gravitational lensing in GRBs \citep{2011AIPC...1358..17} where searches are
carried out for similar temporal and spectral characteristics among several
hundreds of GBM GRB light curves. As a result, morphological GRB classification
attempts have not been successful and the only established division of bursts
into classes with different temporal characteristics is based on their
$T_{90}(T_{50})$ durations defined as the times during which 90\% (50\%) of the
total signal counts (or fluence) are collected \citep{1993ApJ...413L.101K}. The
burst durations when measured in the 50--300 keV energy range, have been found
to distribute bimodally, with over 75\% of the events belonging in the long
class ($>2$\,s).
%

\begin{figure}
\centerline{\includegraphics[width=9.0cm]{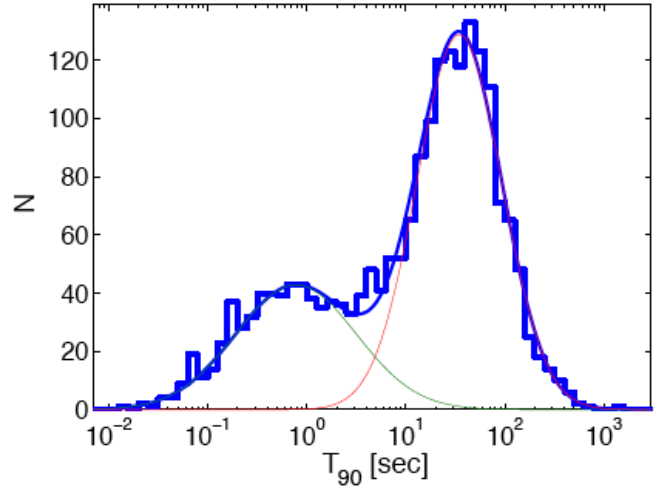}}
\caption{The bimodal duration distribution of GRBs. The blue histogram
represents the distribution of durations for 2041 bursts detected by BATSE. The
histogram is well fit by two log-normal functions (blue and red thin solid
lines) and their sum is shown in thick blue line
\citep{2002A&A...392..791H}.\label{f:T90}}
\end{figure}

\citet{1994MNRAS...271..662} and later \citet{2002A&A...392..791H} showed that
$T_{90}$ of both long and short GRBs follow log-normal distributions separately
(see Fig.~\ref{f:T90}). The distributions peak at $\approx 0.8$~s for short
bursts and at $\approx 32$~s for long bursts. In general, short bursts are
often found to be less luminous by about 3 orders of magnitude leading to the
possibility that excess of long GRBs compared to short ones is most likely an
instrumental selection effect \citep{2007PR...442..166}.

\begin{figure}
\centerline{\includegraphics[width=7.0cm]{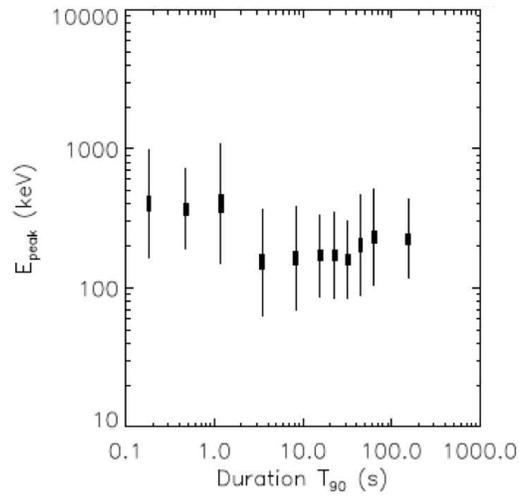}}
\caption{One of the spectral parameters ($E_{\rm peak}$, see \S\ref{ss:cgro}
for details) as a function of burst duration (T90) for spectral fits of a
Comptonised model to BATSE GRBs. The vertical bars show the width of a Gaussian
fit to the parameter distribution within a duration bin. Thick vertical bars
show the error on the mean of each. The average value of $E_{\rm peak}$ for
short bursts is almost a factor of 2 larger to that of long
GRBs.\label{f:T90_hardness}}
\end{figure}

Hardness ratios of short GRBs compared to those of long ones show that the
former are on the average harder. \citet{1993ApJ...413L.101K} combined this
result together with the bimodal duration distribution to suggest that short
and long GRBs are two distinct populations. Fig.~\ref{f:T90_hardness} shows a
plot of the average $E_{\rm peak}$ of a set of BATSE GRBs in different duration
ranges as a function $T_{90}$ clearly showing that on the average the mean
$E_{\rm peak}$ of short GRBs is almost twice as large as that for long GRBs
supporting the dual population hypothesis \citep{2001grba.conf...13}.

\subsection{Spectroscopy of the prompt emission}\label{ss:spec}

Gamma-ray spectroscopy with detectors such as BATSE or later GBM, is not
straightforward like that with optical or X-ray telescopes. These $\gamma$-ray
instruments with high background are actually not measuring photon numbers
directly but counts. The detector response matrices, which allow the conversion
from photon space to count space are usually not invertible. It is then not
possible to measure a photon flux from the measured count flux. The only way to
determine the best fit to a gamma-ray spectrum is to assume a spectral model
and to optimise its parameters by fitting the model to the data following this
procedure:
\begin{itemize}
\item a spectral photon model is assumed;
\item this photon model, with chosen initial parameters, is folded through the
instrument response function to convert photon numbers to counts to be able to
compare the resulting model count spectrum with the real count spectrum
measured by the instrument;
\item the previous step is repeated by varying the model parameters until the
fit convergence criteria is reached.
\end{itemize}
This methodology is a major inconvenience to measuring the real photon spectrum
since the observed spectrum could only be compared with the deconvolved model
dependent spectrum. This makes the identification of the spectral shape more
difficult.

Unlike their light curves, the time-integrated spectra of GRBs do not show the
same extent of diversity. Majority of the BATSE GRB prompt emission spectra in
the keV--MeV energy range were adequately fit with an empirical function called
the Band function \citep{1993ApJ...413..281B}. The Band function, shown below,
consists of two power laws, connected together at the break energy given by
$E_{\rm peak}/(\alpha-\beta)$, with $\alpha$ and $\beta$ the lower and the
higher energy spectral indices respectively:
\[
  N(E) =A \cases{ \displaystyle
   \left( \frac{E}{\rm 100~keV} \right)^\alpha
     \exp\left( -\frac{E(\alpha-\beta)}{E_{\rm peak}}  \right)
   & if $E < E_{\rm peak},$ \cr
   \displaystyle
   \left( \frac{E_{\rm peak}}{\rm 100~keV} \right)^{\alpha-\beta}
     \exp(\beta-\alpha) \left( \frac{E}{\rm 100~keV} \right)^\beta
   & if $E \geq E_{\rm peak},$ \cr}
\]
where $N(E)$ is the differential photon spectrum, $E$ the photon energy in keV
and $A$ is the normalisation constant in photons s$^{-1}$cm$^{-2}$keV$^{-1}$
and $E_{\rm peak}$ corresponds to the maximum of the $\nu F_{\nu}$ (i.e.\ $E^2
N_{\rm e}$) spectrum when $\alpha > -2$ and $\beta < -2$
\citep{1997NCimB...112..11G}.

Although purely empirical, the Band function is usually associated with the
synchrotron emission from electrons that are propagating and accelerated within
the GRB jet \citep{1998ApJ...506L..23P}. However, the values of the Band low
energy power law indices $\alpha$ are inconsistent with the synchrotron slow
and fast cooling scenarios in 20\% of the cases \citep{1998ApJ...506L..23P,
1997ApJ...479L..39C}. For BATSE GRBs, the $E_{\rm peak}$ distribution peaks
around 200 to 300 keV while the $\alpha$ and $\beta$ distributions peak around
$-1$ and $-2$ respectively \citep{2010APJ...721..1329}. As mentioned before,
the short GRBs are usually harder than the long ones with respect to both
$\alpha$ and $E_{\rm peak}$ \citep{2003AIPC...662..248}. A global hard to soft
spectral evolution was reported by \citet{1995ApJ...439..307F} by fitting the
time-resolved spectra to Band functions.

\begin{figure}
\centerline{\includegraphics[width=11.0cm]{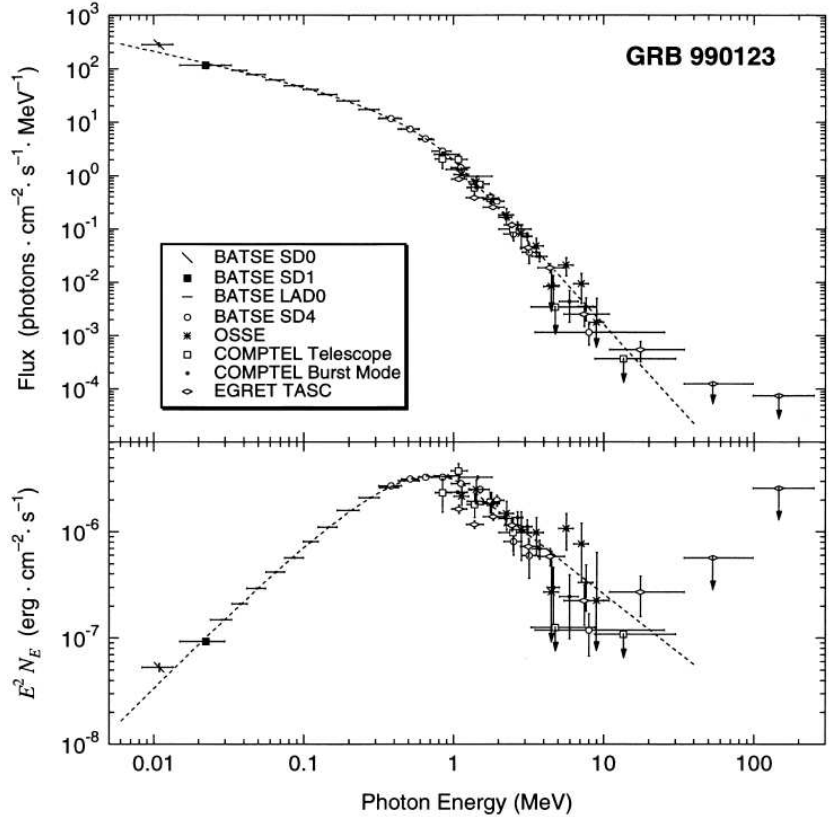}}
\caption{GRB prompt emission spectra in the BATSE Era were usually considered
to be adequately fit with the empirical Band function even on large energy
range in the $\gamma$-ray regime. Here is the case of the famous GRB\,990123
that was simultaneously observed with the 4 instruments on-board CGRO (i.e.\
BATSE, OSSE, Comptel and EGRET) up to several tens of MeV. The top panel shows
the spectrum of this GRB in photon flux units, while the bottom one corresponds
to the $\nu F_{\nu}$ (i.e.\ $E^2N_{\rm e}$) spectrum. This bright GRB has a
high $E_{\rm peak}$ around 1 MeV compared to average GRBs observed with CGRO
\citep{1999ApJ...522..L97}.\label{f:grb990123}}
\end{figure}

\citet{1999ApJ...524..82B} showed that the time-integrated prompt emission
spectrum of the bright GRB\,990123 which was simultaneously observed with BATSE
as well as the three other $\gamma$-ray detectors on board CGRO viz.\ the
Oriented Scintillation Spectrometer Experiment (OSSE;
\citealt{1993ApJS...86..693}), the Compton Telescope (COMPTEL;
\citealt{1989NuPhS...10..121}) and the Energetic Gamma-ray Experiment Telescope
(EGRET; \citealt{1992ITNS...39..993}), could be well fit with a single Band
function across a large energy range from a few tens of keV to about 100 MeV
(see Fig.~\ref{f:grb990123}). \citet{2003Nat...424..749} on the other hand
reported a deviation from the Band function at high energies for one
GRB\,941017, that was detected with both BATSE and EGRET. The $\gamma$-ray
emission excess over the Band function was adequately fit with an additional
high energy power law. A time resolved spectral analysis showed that the
additional power law component stayed constant across the burst while the Band
function evolved as expected, with a global decrease of the emission intensity
together with a softening of the spectrum. This high energy power law component
was found to be inconsistent with the synchrotron shock model.

Delayed high energy $\gamma$-ray emission with respect to the main part of the
prompt emission was reported in several cases \citep{1994ApJ...422..63,
1994Nat...372..652}. A classic example was the case of GRB\,940217 when, among
others, an 18 GeV photon was detected by EGRET, 90 min after the BATSE trigger
time.

In general BATSE bursts exhibited non-thermal spectra in their prompt emission.
However, the fireball model predicts and requires a photospheric emission
simultaneously with the prompt emission \citep{2005ApJ...628..847,
2000ApJ...530..292}. The existence of such a thermal emission component was
claimed at least in some of the BATSE bursts \citep{2003AAP...406..879,
2004ApJ...614..827, 2005ApJ...625L..95, 2010ApJ...709..172}. While
\citet{2003AAP...406..879} tried to fit only a Planck function to the data,
others were using a combination of a black body component and a power law
\citep{2004ApJ...614..827, 2005ApJ...625L..95}.

\section{Observational status: Fermi era}\label{s:Fermi}

\subsection{Fermi Gamma-ray Space Telescope}\label{ss:Fermi_Instr}

The Fermi Gamma-ray Space Telescope (or Fermi for short) -- formerly Gamma-ray
Large Area Space Telescope (GLAST) -- is a space based $\gamma$-ray observatory
launched on 2008 June 11 to explore the gamma-ray sky.  Fermi consists of two
instruments on board viz.\ the Large Area Telescope (LAT) and the Gamma-ray
Burst Monitor (GBM).

The LAT is an imaging high-energy $\gamma$-ray telescope covering the energy
range from $\sim 20$ MeV to more than 300~GeV\footnote{So far spectroscopy only
above 100~MeV is possible while new techniques are in the process of validation
to decrease this threshold at least for spectroscopy of transient sources.}.
The LAT is a pair-conversion telescope with a precision tracker and a
hodoscopic calorimeter consisting of 16 modules in the form of a $4 \times 4$
matrix and a segmented anti-coincidence detector that covers the tracker array.
Each tracker module has a vertical stack of 18 $x,y$ tracking planes,
consisting two layers ($x$ and $y$) of single-sided silicon strip detectors.
The 16 planes at the top of the tracker are inter-leaved with high-$Z$
converter material (Tungsten). Every calorimeter module has 96 CsI(Tl)
crystals, arranged in an 8 layer hodoscopic configuration with a total depth of
8.6 radiation lengths. The aspect ratio of the tracker (height/width) is 0.4
allowing a large field-of-view (2.4 sr). LAT has a programmable GRB trigger and
data acquisition system. The vital specifications of LAT in comparison to those
of EGRET are summarised in Table~\ref{lat_spec} \citep{2009ApJ...697..1071}.

The GBM comprises 12 uncollimated Sodium Iodide (NaI(Tl)) detectors operating
over the 8~keV to 1\,MeV range, and 2 Bismuth Germanate (Bi$_4$Ge$_3$O$_{12}$
or BGO for short) detectors operating over the 150~keV to 40~MeV range. The
axes of the NaI detectors are oriented such that they have the uniform view of
the entire unocculted sky and the positions of GRBs can be derived from the
measured relative counting rates, a technique previously employed by BATSE.
Although GBM is smaller and slightly less sensitive than its predecessor BATSE,
its extended energy range, from 8 keV to 40 MeV, combined with an improved data
format, makes it an unprecedented all-sky instrument for time-integrated and
fine-time resolved spectroscopy of transient sources like GRBs, Terrestrial
Gamma-ray Flashes (TGFs), Soft Gamma Repeaters (SGRs), Solar Flares etc. The
GBM flight software triggers on a GRB (or any fast transient) with a peak flux
above 0.75 photons cm$^{-2}$s$^{-1}$ and locates it in the sky within an error
of $\sim 15^\circ$ and transmits the information through Gamma-ray Coordinates
Network (GCN\footnote{\texttt{http://gcn.gsfc.nasa.gov/}}) within the first 8
seconds of a trigger to initiate follow-up observations with other instruments.
Updates on location with an accuracy of $\sim 5^\circ$ are sent within next
couple of seconds. More details on GBM detectors and data types may be found in
\citet{2009ApJ...702..791}. GBM also provides near real-time on-board
burst-localisation to the Fermi spacecraft to repoint the LAT in direction of
the source in order to increase its sensitivity to the burst and to detect
possible high energy delayed emission from the source.

\begin{table}
\caption{LAT Specifications and Performance Compared with
EGRET.}\label{lat_spec}
\medskip
\centering
\begin{tabular}{crr}\hline
Quantity                          & LAT (Minimum Spec.)         & EGRET \\\hline
Energy Range                      & 20~MeV -- 300~GeV           & 20 MeV -- 30 GeV      \\
Peak Effective Area$^{\rm a}$     & $>8000$ cm$^2$              & 1500 cm$^2$           \\
Field of View                     & $>2$ sr                     & 0.5 sr                \\
Angular Resolution$^{\rm b}$      & $<3.5^\circ$ (100~MeV)      & $5.8^\circ$ (100 MeV) \\
                                  & $<0.15^\circ$ ($>10$~GeV)   &                       \\
Energy Resolution$^{\rm c}$       & $<10$\%                     & 10\%                  \\
Dead-time per Event                & $<100$~$\muup$s             & 100~ms                \\
Source Location Determination$^{\rm d}$ & $<0.5'$	              & $15'$                 \\
Point Source Sensitivity$^{\rm e}$ & $<6 \times 10^{-9}$ cm$^{-2}$s$^{-1}$ &
                                          $\sim 10^{-7}$ cm$^{-2}$s$^{-1}$              \\\hline
\end{tabular}\\[5pt]
\begin{minipage}{11.5cm}
\small Notes: (a) after background rejection; (b) single photon, 68\%
containment, on-axis (c) 1-$\sigma$, on-axis; (d) 1-$\sigma$ radius, flux
$10^{-7}$ cm$^{-2}$s$^{-1}$ ($> 100$~MeV); (e) $>100$~MeV, at high $|b|$, for
exposure of one-year all sky survey, photon spectral index $-2$.
\end{minipage}
\end{table}

\subsection{Synergy between Fermi and the other instruments}

Currently, GBM is the best instrument to study GRB prompt emission in the
keV--MeV energy range where the prompt emission is the most intense. With the
LAT, our knowledge of the prompt emission at high energy is extended up to
several tens of GeV. Even though Fermi is an outstanding observatory for
studying the prompt emission, the poor location capabilities of GBM (several
degrees) is a handicap for follow-up observations. To perform spectral analysis
with GBM, detector response matrices must be generated based on the source
location. Using an inaccurate localisation can lead to wrong responses if, for
instance, a part of the spacecraft, such as a radiator, blocks one or more GBM
detectors from the source. Accurate localisation is also necessary to repoint
the ground based optical and VHE $\gamma$-ray telescopes (see \S\ref{ss:vhe}
for details) that have narrow field of view. Without an accurate localisation
of the source, we cannot get counter part signature at other wavelengths, nor
host galaxy detection, nor redshifts. These observations are crucial to
understand GRBs. Hence a good synergy of Fermi with other instruments is
required. Obviously, the LAT can locate a GRB accurately enough to perform
follow up observations with other instruments, but only a few GRBs (about 18)
have been detected by the LAT (during first 2 years of operation) over a total
$\sim 500$ GBM bursts, and the locations were obtained from ground analysis
several hours after the burst trigger, which is late to initiate a successful
follow-up observations.

IPN and Swift can provide good localisations. The IPN location is too delayed
to be useful for follow-up observations as mentioned before. Swift detected
$\sim 14$\% of GBM GRBs during the first 2 years of operation, of which $\sim
12$\% triggered the Swift BAT and the rest were discovered after ground
analysis of BAT data. $\sim 1$\% were located by the Swift XRT based on trigger
from another instrument \citep{2011ApJ}.

Detection of GBM GRBs with other gamma-ray instruments is also important to
inter-calibr\-ate the detectors. Table \ref{coic_missions} lists the fraction of
GBM GRBs detected simultaneously with other $\gamma$-ray instruments during the
first 2 years operation.

\begin{table}
\caption{Fraction of GBM GRBs also detected with other gamma-ray instruments
during the first 2 years till July 15, 2010.}\label{coic_missions}
\medskip
\centering
\def\hpad{\kern20pt}
\begin{tabular}{crr}\hline
Instrument                    & \# of Observed GRBs       & Fraction of \\
                              &  simultaneously with GBM  & GBM GRBs (\%)\\\hline
Konus (WIND)                  & 146 \hpad    & $\sim 30$ \hpad  \\
Integral (Burst Alert System) &   6 \hpad    &  $\sim 1$ \hpad  \\
Integral (SPI-ACS)            & 213 \hpad    & $\sim 43$ \hpad  \\
Swift                         &  68 \hpad    & $\sim 14$ \hpad  \\
Super AGILE                   &   5 \hpad    &  $\sim 1$ \hpad  \\
Fermi (LAT)                   &  18 \hpad    &  $\sim 4$ \hpad  \\
RHESSI                        &  33 \hpad    &  $\sim 7$ \hpad  \\
MAXI                          &   4 \hpad    &  $\sim 1$ \hpad  \\
Suzaku (WAM)                  & 200 \hpad    & $\sim 41$ \hpad  \\\hline
\end{tabular}
\end{table}

\subsection{GRB temporal results}\label{ss:temp_res}

\subsubsection{Energy dependence of burst durations}\label{sss:t90}

\begin{figure}
\centerline{\includegraphics[width=12.0cm]{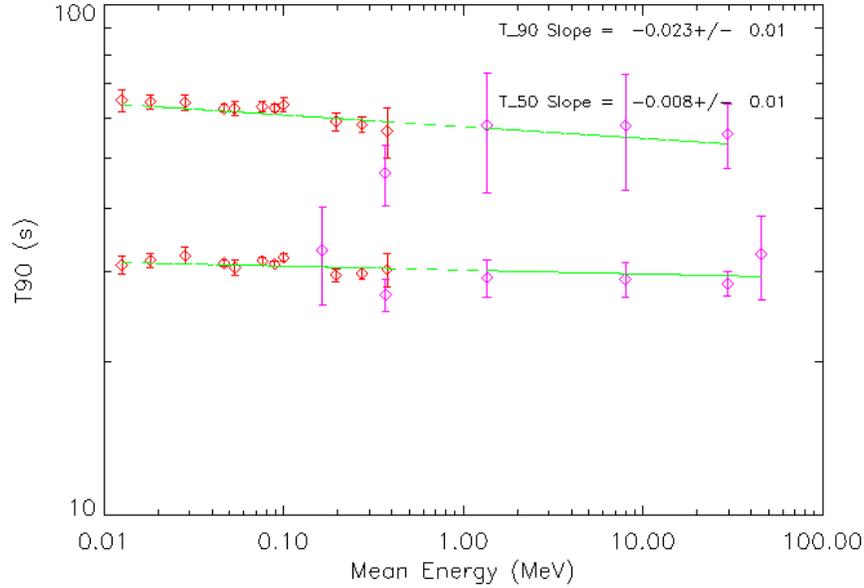}}
\caption{The upper plot shows the variation of $T_{90}$ as a function of mean
energy of the $\gamma$-rays. The straight line shows a power law fit to the data
and the slope is indicated. Similarly the lower plot corresponds to $T_{50}$
values of the same burst. The $T_{90}$ and $T_{50}$ values evolve differently
and not consistent with the average behaviour.\label{f:en_T90}}
\end{figure}

The observer frame GRB durations (as described in \S\ref{ss:t90}) measured are
expected to suffer from relativistic effects. It was realised that the burst
intensity could add unknown systematic errors in the estimation of durations as
well \citep{1997ApJ...490...79}. Often weaker GRBs have large errors on their
estimated durations. See the recently published GBM 2-year burst catalog for a
comprehensive summary of GBM GRB durations and fluences \citep{2011ApJ}.
\citet{2011AAP...531..20} studied the GRB rest frame distributions of 32 GBM
GRBs with known redshifts. They find that the $T_{90}$ distribution in the GRB
rest frame peaks at $\approx~$10\,s. Another important parameter that affects
the estimation of $T_{90}$ is the energy range \citep{1996AIPC...384..87,
2011ApJ...733...97}. It has been found that on the average the burst durations
fall with increasing energy as a power law with an exponent of $\sim 0.4$.
However for a given burst the duration could evolve differently with energy.
The $T_{90}$ and $T_{50}$ values for the GBM detected burst GRB\,080916C  are
estimated in different energy bands using the GRB light curves summed over 4
bright NaI and 2 BGO detectors separately. The $T_{90}$($T_{50}$) are plotted
as a function of average energy and shown in Fig.~\ref{f:en_T90}. The BGO
points are shown in purple. The slopes of the power law fits to the durations
are also indicated in green. The values of the slopes (shown on the top right)
are much shallower than $-0.4$.

\subsubsection{GRB spectral lags}\label{sss:lag}

Temporal delays in the arrival of low-energy photons relative to that of
high-energy photons, called spectral lags, are well known in GRBs. A general
convention is positive spectral lag corresponds to an earlier arrival time of
the higher energy photons. Spectral lags are traditionally measured between a
pair of light curves of a GRB in 2 energy bands (25--50 and 100--300 keV for
BATSE bursts). The standard method adapted is to compute the cross-correlation
function (CCF) between the two light curves. The time lag at which the CCF
peaks is defined as spectral lag. Generally the spectral lags of long GRBs are
positive, a small fraction of them have been shown to exhibit zero or negative
lags \citep{2002Astro-ph...0206402, 2005ApJ...619..983}. One of the proposed
explanations for the observed spectral lag is the spectral evolution during the
prompt phase of the GRB \citep{1998ApJ...501..L157, 2003ApJ...594..385,
2005AAP...429..869}. Due to cooling effects, $E_{\rm peak}$ moves to a lower
energy channel after some characteristic time. When the $E_{\rm peak}$ moves
from a higher energy band to a lower energy band, the temporal peak of the
light curve also moves from a higher energy band to a lower one, which results
in the observed spectral lag. In a recent study \citet{2011AN...332..92}
suggest that spectral evolution can be invoked to explain both positive and
negative spectral lags. Hard-to-soft evolution of the spectrum produces
positive spectral lags while soft-to-hard evolution would lead to negative
lags.

Regardless of its physical origin, the spectral lags show some interesting
correlations with other measured parameters. Based on six GRBs with known
redshifts, \citet{2000ApJ...534..248} found an anti-correlation between the
spectral lag and the isotropic peak luminosity ($L_{\rm iso}$). Further
evidence for this correlation was provided by many others
\citep{2002ApJ...579..386, 2006Nat...444..1044, 2007ApJ...660..16,
2008ApJ...677..L81}. More recently \citet{2011MNRAS...410} using 43 Swift long
bursts with known redshifts showed that spectral lag and $L_{\rm iso}$ exhibit
a higher degree of correlation in the GRB source frame. Many authors have
provided possible explanations of the physical cause of lag luminosity
relation. Observed spectral lags, as well as peak luminosity, naturally have a
strong dependence on the Doppler factor of the outflow (a function of the
Lorentz factor and the direction of motion with respect to the observer). If
indeed the Doppler factor is the dominant parameter among GRBs, then a relation
between spectral lags or variability and luminosity is expected
\citep{2002ApJ...569..682}. \citet{2000ApJ...544..L115} on the other hand
argues that the anti-correlation is due to the variations in the line-of-sight
velocity of various GRBs. \citet{2001ApJ...554..L163} suggest that the relation
is a result of variations of the off-axis angle when viewing a narrow jet.
Alternatively, \citet{2004ApJ...602..306} invokes a rapid radiation cooling
effect to explain the correlation.

It has been shown that short GRBs have either small or negligible lags
\citep{2006ApJ...643..266, 2006MNRAS...373..729}. The average spectral lag as
measured for 32 short duration GRBs detected by GBM lies in the range $\pm
32$~ms \citep{2011AIPC...1358..183}. According to the lag--$L_{\rm iso}$
relation, these small lag values imply that short bursts are highly luminous.
However, based on the redshift measurements of their host galaxies it has been
shown that short GRBs are generally less luminous than long bursts. Hence short
bursts seem to not follow the lag-luminosity relation
\citep{2006Nat...444..1044}. The spectral lag has often been used to identify a
short burst from the long ones.

An anti-correlation has also been derived between the GRB spectral lag and the
jet opening angle \citep{2002ApJ...569..682}. This relation, when coupled with
the above relation between spectral lag and $L_{\rm iso}$ is another
manifestation of the anti-correlation between the isotropic luminosity and the
jet opening angle, which is the direct consequence of the standard energy
reservoir relation \citep{2001ApJ...562..L55}.

\begin{figure}
\centerline{\includegraphics[width=11cm]{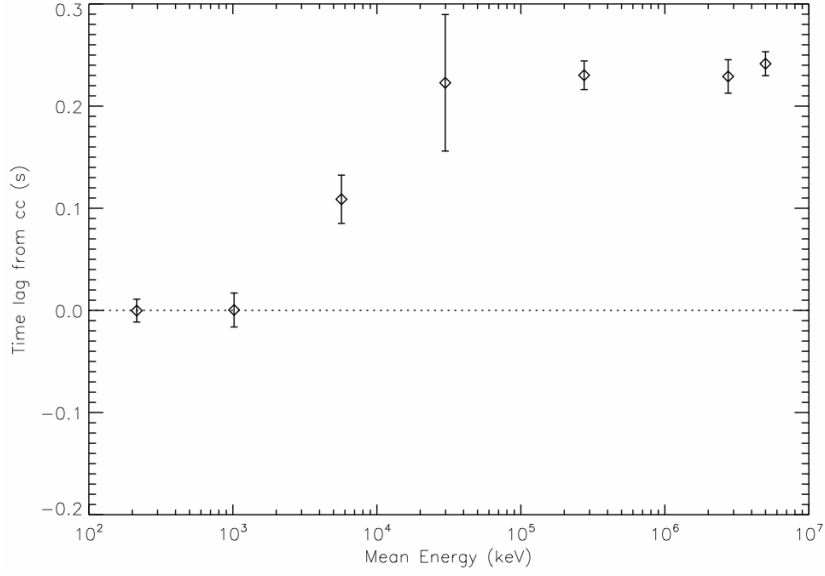}}
\caption{A plot of the spectral lag for the short burst GRB\,090510 as a
function of mean $\gamma$-ray energy. The light curves in the range 8--1000~keV
are from the GBM NaI detectors while those in the energy range 1--40~MeV are
from the BGO detectors and from the Large Area Telescope (LAT) for the higher
energy light curves.\label{f:lag}}
\end{figure}

Since the GBM has a large energy bandwidth between the NaI and the BGO
detectors (8~keV--40~MeV) it has been possible to study the evolution of GRB
spectral lags with energy. Such a study led to new discovery that in some GRBs
the spectral lag changed sign at higher energies. For example, GRB\,090510
shows no significant spectral lag measured between lowest energy light curve
(8--20~keV) and higher energy bands up to an average energy of about an MeV and
shows a negative lag at a few MeV and beyond. It saturated at constant value of
about 250\,ms and remained constant there after \citep{2010ApJ...725..225G,
2011AIPC...1358..183}. This could be interpreted as due to the differences in
the production processes of low and higher energy $\gamma$-rays (see Fig.~\ref{f:lag}).

\subsubsection{GRB light curve decomposition studies}\label{sss:lc_decom}

Most GRB light curves are highly variable with variability time scale
significantly smaller than their overall duration. According to the internal
shock model \citep{2003AIP...662..3}, the central engine generates relativistic
shells with highly non-uniform distribution of Lorentz factors and the pulses
are formed by the collision between a rapidly moving shell with a slower shell
as mentioned before. Thus in principle the variability of the GRB light curves
may directly correspond to the activity of their central engines
\citep{2003MNRAS...587..592, 2002ApJ...572..L139}. Hence the studies of pulse
properties are important to determine whether GRB sources require engines that
are long lasting or impulsive \citep{2004ApJ...614..284}. Their temporal
deconvolution can reveal potential differences in the properties of the central
engines in the two populations of GRBs  which are believed to originate from
the deaths of massive stars (long) and from mergers of compact objects (short).
Several authors have studied the deconvolution of GRB light curves into their
constituents, and have shown that in general, these are discrete, often
overlapping pulses with durations ranging from a few milliseconds to several
seconds and almost always asymmetric shapes, with faster rises than decays
\citep{1996ApJ...459..393, 2000AIPC...526..215, 2011ApJ...741,
2011ApJ...740..104}. These highly varied GRB temporal profiles are suggestive
of a stochastic process origin.

In recent years, pulse properties have provided increasingly valuable
constraints on the physics responsible for GRB prompt emission. These
properties have included (1) temporal asymmetry characterised by longer decay
than rise times, (2) hard-to-soft spectral evolution, (3) broadening at lower
energies and (4) decreasing pulse height with time since trigger for those GRBs
which show hard-to-soft spectral evolution \citep{1996ApJ...459..393,
2002Astro-ph...0206402}.

\begin{figure}
\centerline{\includegraphics[width=11cm]{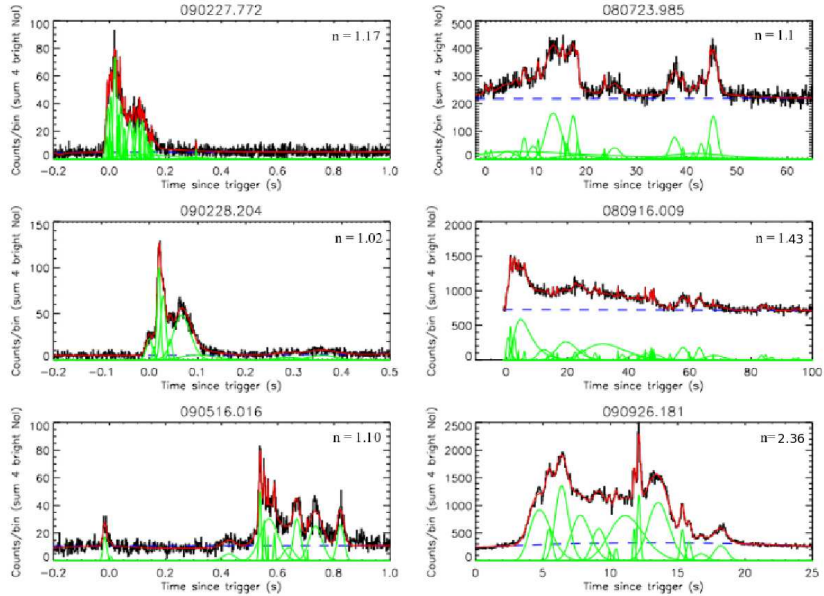}}
\caption{A sample fits two the light curves of 3 short (left column) and 3 long
GRBs detected by GBM. Also indicated on the top right of each panel a number
equivalent to $\chi^2$ showing the goodness of fit in each case.\label{f:fits}}
\end{figure}

Using pulse data contained in individual energy channels, it has been
demonstrated that not only the pulses have longer durations at lower energies,
but also peak later at the lower energies \citep{2009AIPC...1133..379,
2008ApJ...677..L81}.

\begin{figure}
\centerline{\includegraphics[width=6.2cm]{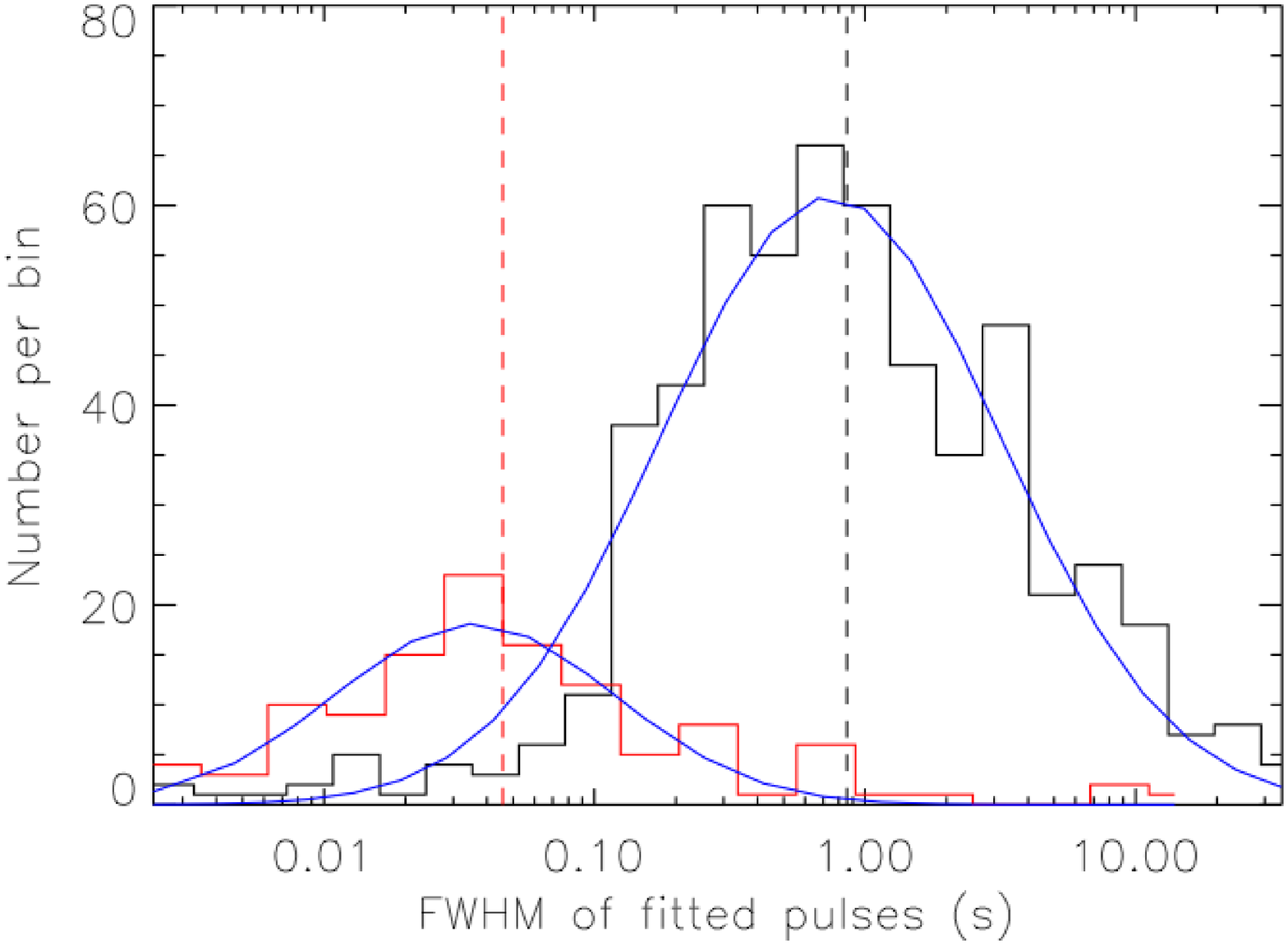}
            \includegraphics[width=6.8cm]{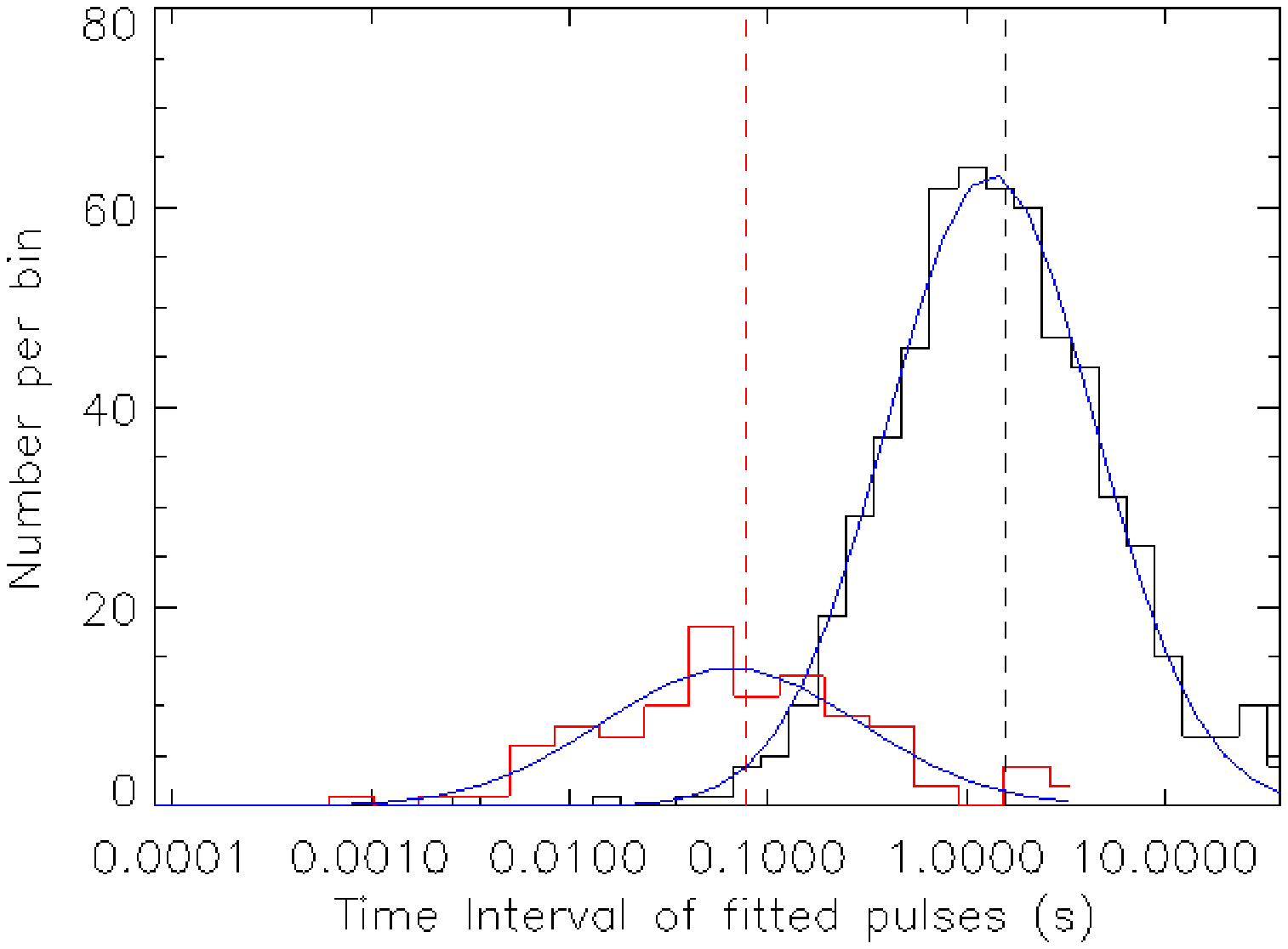}}
\caption{(left) Distributions of the pulse widths (FWHM) for long (histogram
shown in black) and short bursts (histogram shown in red). A log-normal function
is fitted to each of the distributions. The mean values of FWHM (from the fit)
for long and short bursts  are 0.89\,s and 0.055\,s and the standard deviations
are 5.2\,s and 4.6\,s respectively. The vertical dashed lines are the median
values of FWHM for each class of GRBs. (right) Distributions of the time
intervals between successive pulses ($\Delta t$) for long (histogram shown in
black) and short (histogram shown in red) bursts. A log-normal function is
fitted to each of the distributions. The mean values of $\Delta t$ (from the
fit) for long and short bursts are 1.53 s and 0.076\,s and the standard
deviations are 3.6 and 5.1 respectively. The vertical dashed lines indicate the
median values of the time intervals between successive pulses for each class of
GRBs.\label{f:many}}
\end{figure}

Fig.~\ref{f:fits} shows a sample of 3 short bursts (left column) and 3 long
bursts (right column) whose light curves are deconvolved by fitting log-normal
shapes (shown in green at the bottom of each plot) to each pulse in the light
curve and then superposed over a quadratic background (shown as dashed line).
The resulting fit shown as continuous line in red fits the actual light curve
(black histogram) very well. Also shown are the reduced $\chi^2$ values on top
right corner in each plot. \citet{2011ApJ...741} also find fundamental
differences in the pulse properties of long and short duration bursts.
Fig.~\ref{f:many} (left) shows the distribution of pulse widths and (right) interval
between successive pulses for long and short bursts. It is found that both the
pulse widths and pulse intervals follow log-normal functions as shown by the
fits separately for long and short bursts. The observed separation of the
distributions for the two classes of bursts demonstrates that the short burst
light curves consist of more closely packed narrower pulses while long burst
light curves consist of broader pulses with larger separations between them.
These conclusions support internal shock model for GRBs.

\subsubsection{GRB timescales}\label{sss:timescale}

According to the internal shock model, the GRB pulses are formed by the
collisions among relativistic shells ejected by the central engine with a
distribution of Lorentz factors (${\Gamma}$). A GRB pulse shape depends on
three time scales. The hydrodynamic time scale, $t_{\rm dyn}$ (that determines
the pulse rise time), the angular spreading time scale, $t_{\rm ang}$ (that
determines the pulse decay time), and the cooling time scale, $t_{\rm rad}$
(which is usually much shorter than the other two and can be ignored)
\citep{1997ApJ...490..92, 1997ApJ...490..663, 1996ApJ...473..998}.

\begin{figure}
\centerline{\includegraphics[width=9.0cm]{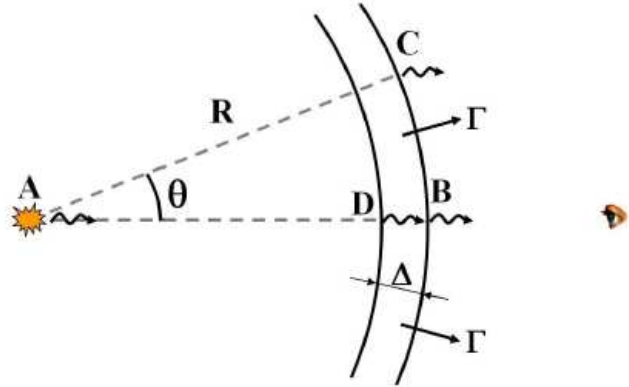}}
\caption{ A cartoon of a spherical relativistic shell of width $\Delta$
expanding with a Lorentz factor $\Gamma$ Four photons are emitted as indicated.
These photons set three typical time scales. Photon `A' is emitted from the
origin at the time of the explosion. Photons `B',`C' and `D' are all emitted
when the shell front is at radius R. `B' is emitted from the front of the
relativistic shell on the line-of-sight. `C' is emitted from the front of the
shell at an angle $\theta$. `D' is emitted from the rear of the shell on the
line-of-sight.\label{f:shell}}
\end{figure}

In Fig.~\ref{f:shell}, a spherical shell of width $\Delta$ is expanding with
a relativistic Lorentz Factor $\Gamma \gg 1$ and emits a photon at radius $r =
0$ from A, at the time of explosion and at $r = R$ from points B, C and D. The
time interval between two photons emitted by the shell from points A and B is
called the line of sight time, $t_{\rm los}$ and is given by:
\[
  t_{\rm los} = \frac{R}{2c{\Gamma}^2}.
\]
Similarly the angular time scale, $t_{\rm ang}$, i.e.\ the interval between the
two photons emitted by the shell from points B and C also happens to be equal
to $t_{\rm los}$. Similarly the time interval between the two photons emitted
from D and B is given by \citep{1997ApJ...485..270}
\[
  t_{\rm rad} = \frac{\Delta}{2c{\Gamma}^2}.
\]
In the internal dissipation models the burst duration is related to the
duration of the engine activity whereas the variability time scale $\Delta t$
could reflect either the variability of the flow or the irregularities in the
radial structure of the shell. The irregularities that it encounters will be
spread on a time $t_{\rm ang}$. So $\Delta t \geq t_{\rm ang}$. If the
thickness of the shell is $\Delta$ in the source rest frame, the rest frame
duration of the burst must be longer than $\Delta/c$. In the case of external
dissipation models of baryonic outflow energy can be released on an engine
dynamical time scale. The duration of the burst in this model is $R_\gamma/10 c
\Gamma^2$ where $R_\gamma$ is the radius at which blast wave associated with
the fireball becomes radiative \citep{1998ApJ...492..677}. The variability time
scale in this case is determined by the size of the density fluctuations in the
external medium.

Variability time scale is an important parameter which is related to the
characteristic timescales of emission process mentioned above. The FWHM of the
pulses in a deconvolved light curve may be considered as a measure of
variability time scale at any time during the prompt emission. Hence this
technique of deconvolving the GRB light curve into pulses enables one to trace
the evolution of emission time scale during the prompt emission while a median
or a mean FWHM may be considered as a typical variability time scale for a GRB.

The radius in the source frame at which prompt emission takes place can be
estimated from the variability time scale as follows:
\[
  R_{\rm e} \approx \frac{2\gamma_{\rm e} ^2 c t_{\rm var}}{1+z}
\]
where $t_{\rm var}$ is the GRB light curve variability time scale
\citep{2004ApJ...614..284}. Using the range of pulse widths required to
deconvolve the light curve one can estimate the range of radii where the
internal shocks are operating resulting in the observed emission.

\subsection{GRB spectral analysis results}\label{ss:spec_ana}

In the Fermi Era, GRB prompt emission spectra observed with GBM are still
usually adequately fit with the Band function from 8 keV to 40 MeV
\citep{2011APJSG, 2011ApJ...733...97} with spectral parameters mostly consistent with 
what was reported from BATSE data \citep[][Goldstein et al., in preparation]{2011MNRAS.415.3153N,2011ApJ...733...97}. However, due to
the superior sensitivity of the LAT compared to EGRET, more GRBs have been
observed at high energy above 100 MeV. At the current date, GBM detected 800
GRBs \citep{2011ApJ} and the LAT detected 28 GRBs \citep[Abdo et al., in preparation, ][for a smaller sample]{2011ApJ...730..141Z}. 
20 of the LAT GRBs have a significant emission above 100 MeV. The others have a low
significance above this energy threshold, but they exhibit a clear signal below
(in the 20--100~MeV range).

While in some GRBs, the low energy emission observed in GBM and the high energy
emission detected by the LAT start simultaneously, in others the emission onset
above 100 MeV is delayed compared to the low energy observed in GBM. However,
when started, the structures of the $>$100 MeV light curves match with the ones
observed at lower energy with GBM (see \S\ref{sss:he_delay} for more details).

\begin{figure}
\centerline{\includegraphics[width=11.0cm]{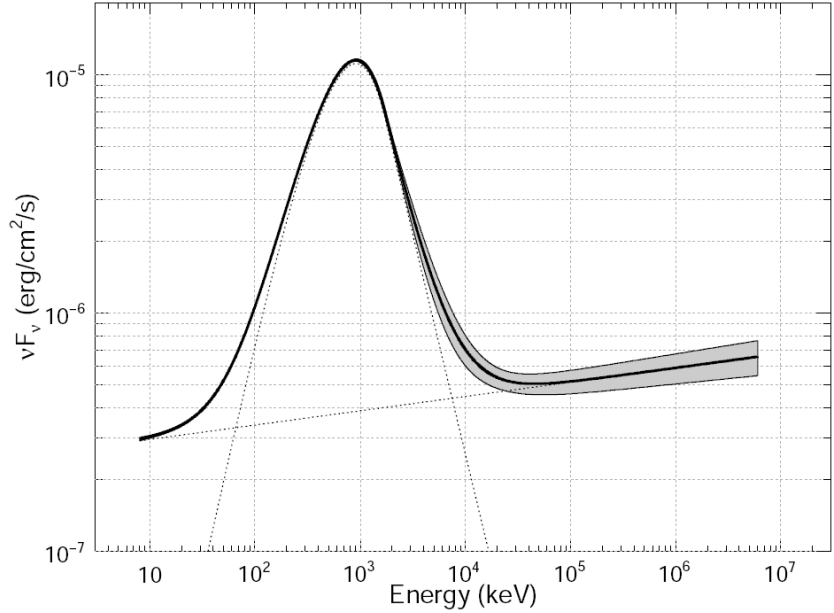}}
\caption{$\nu F_{\nu}$ spectrum of GRB\,090902B as observed with the two
instruments on-board Fermi, GBM and LAT.  The Band function is insufficient to
fit the prompt emission spectrum which requires an additional power law. The
additional power law is more intense than the Band function below few tens of
keV and above several MeV. The thin dashed lines represents the two components
used to fit the spectrum (i.e.\ Band function and power law), and the thick
solid one corresponds to the sum of the two components. The gray butterfly
represents the 1 sigma confidence contour taking into account the covariance
matrix resulting from the fit.\label{f:grb090902}}
\end{figure}

\begin{figure}
\centerline{\includegraphics[width=11.0cm]{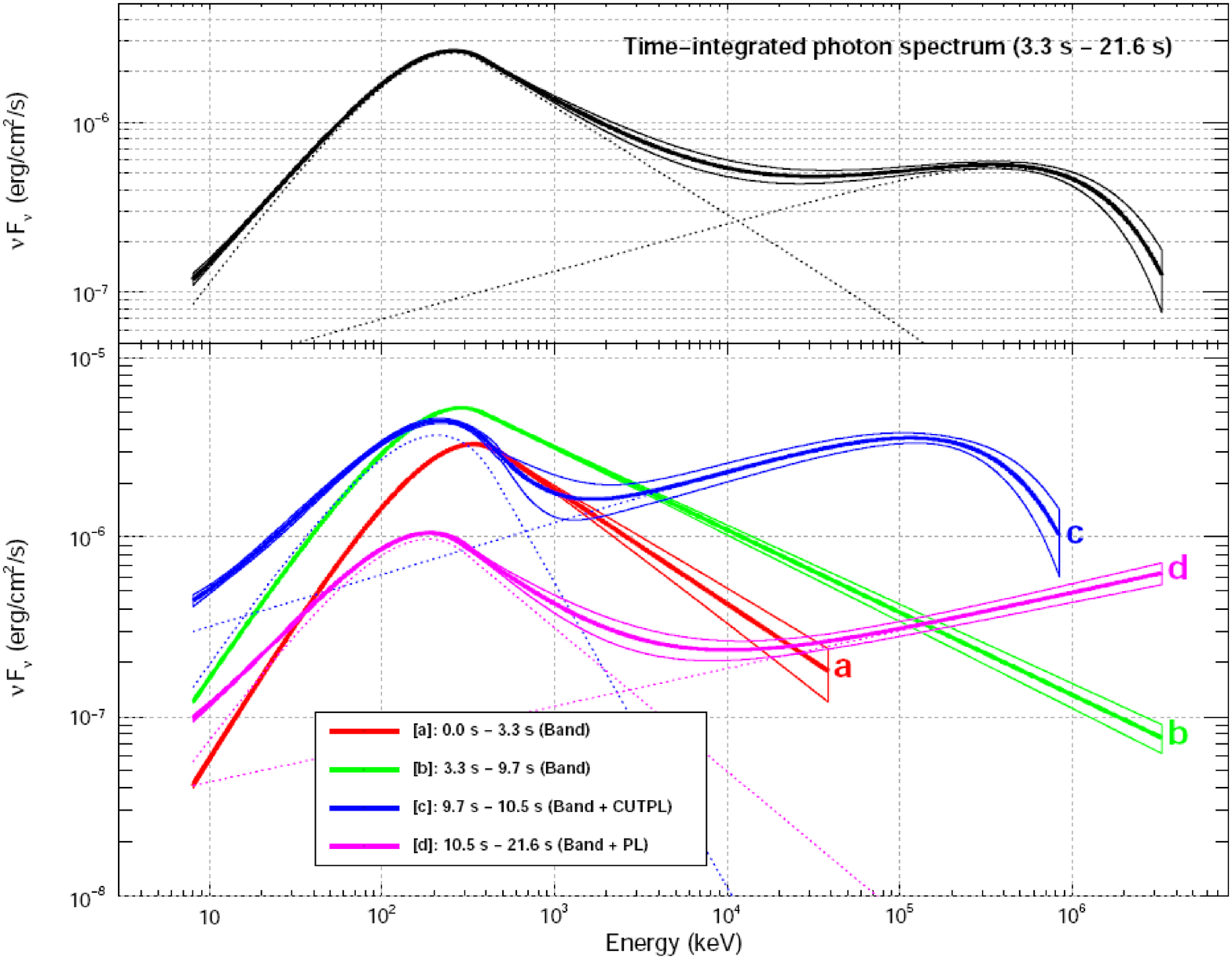}}
\caption{Top panel: $\nu F_{\nu}$ time integrated spectrum of GRB\,090926A. As
for GRB\,090902B (see Fig.~\ref{f:grb090902}), an additional power law is
required together with the traditional Band function. However, in this case, a
break can be measured in the additional power law around tens of MeV. Bottom
panel: Time resolved $\nu F_{\nu}$ spectra of GRB\,090926A. Initially, the
prompt emission spectrum is well fit with a single Band function. After few
second, the additional power law kicks off. In both the panels, the thin dashed
lines represent the two components used to fit the spectrum (i.e.\ Band or Band
function and power law), and the thick solid lines correspond to the sum of the
components. The butterflies represent the 1 sigma confidence contour taking
into account the covariance matrix resulting from the fits.\label{f:grb090926}}
\end{figure}

While some joint GBM and LAT spectral fits are consistent with a Band function
\citep[Abdo et al., in preparation]{2009Sci...323.1688, 2009ApJ...707..580, 2010ApJ...712..558A}, 
some exhibit strong deviations, which are well fit with an additional power law
in addition to the Band function as reported in \citet{2003Nat...424..749} from
CGRO data. Such deviation have been reported several times in both, long
\citep{2009ApJ...706L.138A, 2011ApJ...729..114A} and short Fermi GRBs
\citep{2010ApJ...716.1178A}, and the additional power law usually overpowers
the Band function below few tens of keV and above few tens of MeV as seen in
figures \ref{f:grb090902} and \ref{f:grb090926}. This additional power law,
with an index value, which usually ranges from $-1.5$ to $-1.9$ (harder than
the $\beta$ parameter of the Band function) challenges both leptonic and
hadronic models \citep{2009ApJ...705L.191A}. For instance, leptonic models such
as inverse-Compton (IC) or synchrotron self Compton (SSC) emission can
naturally explain the high energy power law. However, they do not support the
delayed onset of high energy component nor the low energy power law excess.
Hadronic models, through pair cascade or proton synchrotron models
\citep{2009ApJ...705L.191A, 2009ApJ...691L..37R} have difficulties to reproduce
the correlated variability at low and high energies. In addition, proton
synchrotron emission requires very large magnetic fields. However, they can
explain the delayed onset of high energy emission since the time to accelerate
protons or to develop cascades is longer than that for electrons. In hadronic
models, synchrotron emission from secondary electron-positron pairs produced
via photon hadron interactions can naturally explain the power law at low
energy. \citet{2009MNRAS.400L..75K, 2010A&A...510L...7G} also suggested that
the additional power law and the delayed high energy emission could originate
from an early afterglow and produce by electron-positron synchrotron emission
from the external shock. However, the short variability time scale remains a
problem.

In some cases, the additional power law does not extend at high energy but can
still be fit in addition to the Band function from GBM data alone
\citep{2010ApJ...725..225G}. For GRB\,090926A \citep{2011ApJ...729..114A}, a
cutoff was required around 1.4\,GeV in the additional power law, and this
additional component was clearly associated with a very bright peak in the
prompt emission light curve which dominates the rest of the emission in all
energy bands from 8 keV to few tens of MeV (see Fig.~\ref{f:grb090926}).
\citet{2011MNRAS...415..3693, 2010ApJ...709..172} reported the fit of bright
Fermi GRBs with a combination of a modified black body, called multi colour
black body  \citep{2011ApJ...732...49P}, and an additional power law using fine time interval.
Possible broadening of the planck function due to sub-photospheric energy dissipation are 
also discussed by \citet{2006ApJ...642..995P,2010ApJ...725.1137L,2010MNRAS.407.1033B,2011MNRAS.415.1663T}

\begin{figure}
\centerline{\includegraphics[width=11.0cm]{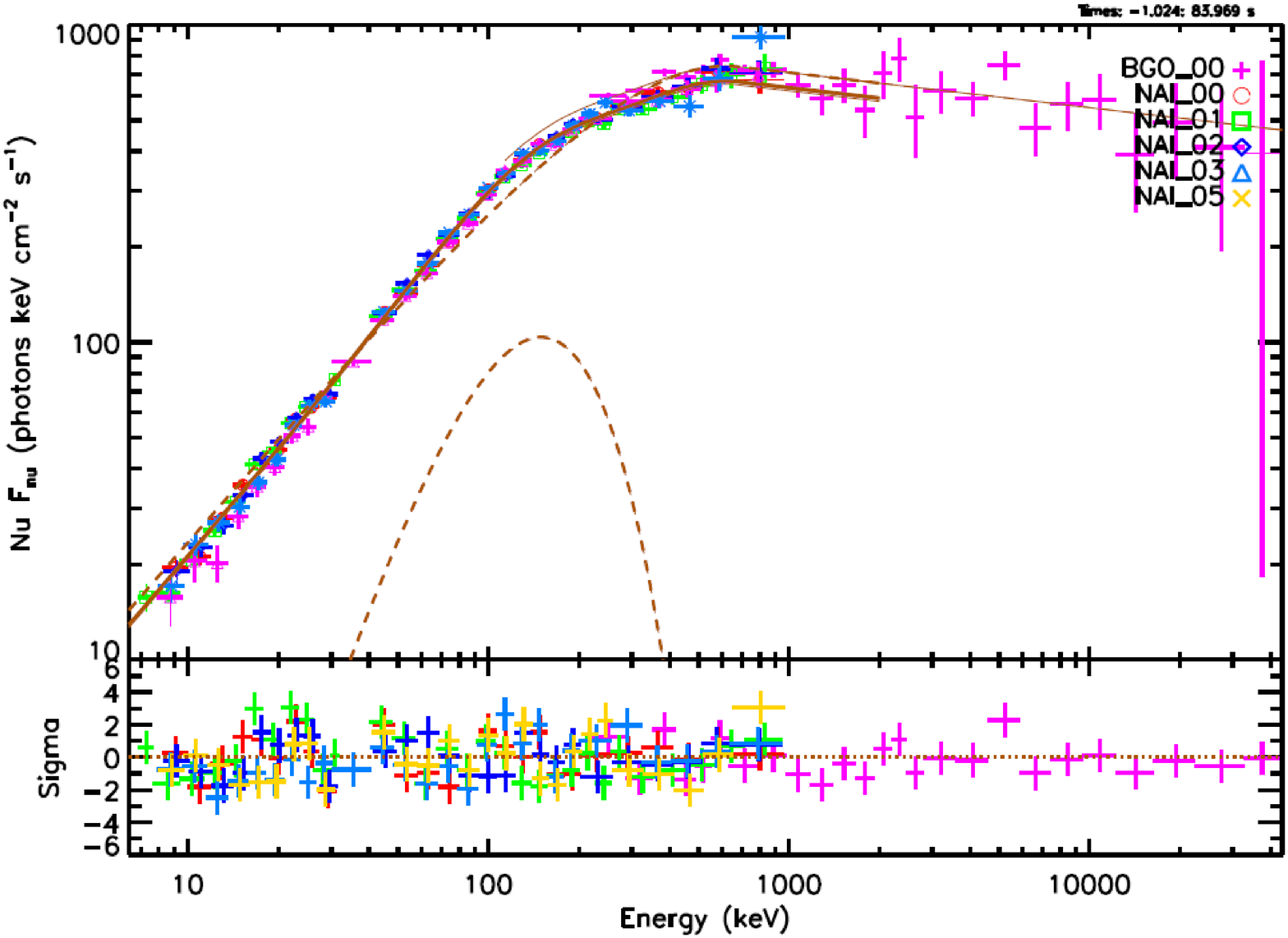}}
\caption{Time-integrated $\nu F_{\nu}$ spectrum of GRB\,100724B as observed
with GBM and fit with a Band function and an additional black body component.
The thin dashed lines represented the two components used to fit the spectrum
(i.e.\ Band function and black body), and the solid line corresponds to the sum
of the two components. The points plotted on the top panel are model dependent
and results from the fit of the real count spectrum with a Band function and a
black body component. Bottom Panel: residual of the fit when the real count
spectrum is fit with a Band function and an additional black body component. A
significant black body component is detected, in addition to the traditional
Band function. The detection of such a component reinforced the fireball model
which predict such a photospheric emission which has never been clearly
detected before. The relatively low intensity of the black body component
compared to the Band function and the temperature of this component (around
$\sim 30$ keV), imply that the energy reservoir cannot be of purely internal
origin, but the outflow has to be highly magnetised close to the
source.\label{f:grb100724}}
\end{figure}

 More recently, \citet{2011ApJ...727..L33} reported for the first time a significant
improvement when fitting the time integrated spectrum of GR\,100724B with a
about 30 keV black body component in addition to the standard Band function
(see Fig.~\ref{f:grb100724}). The evolution of the thermal and the non
thermal components were also followed across the prompt emission using
time-resolved spectroscopy and the black body component was consistent with a
possible cooling. The simultaneous identification of a thermal component in
addition to the traditional non thermal one, is a major step to validate such a
result. This black body component was interpreted as the jet photospheric
emission, and its characteristics implied that the energy
released by the central engine could not be of pure internal origin, but that
the outflow had to be highly magnetized close to the source. Prompt emission
spectra best fit with a combination of a Band function, a black body component
and an additional power law have been observed in several Fermi GRBs \citep[Guiriec
et al., in preparation, ][]{2011AIPC.1358...33G}.

However, lack of detection of thermal emission  in most of the bursts plays against the fireball scenario. As a result some authors focused on magnetically dominated models \citep{2009ApJ...700L..65Z,2011ApJ...726...90Z}.

\citet{2010ApJ...725..225G} reported that the light curves observed at low and
high energies in the GBM data where not matching, the highest energy ones
exhibiting more numerous sharper structures with $E_{\rm peak}$ tracking these
high energy light curves. This suggests the existence of multiple spectral
components which are dominating in different region of the spectrum. The high
energy emission observed in the LAT lasts usually longer than the prompt
emission detected in GBM, and a long lasting high energy emission observed
above 100 MeV has been detected up to several ks in some cases
(Abdo et al., in preparation).  However, it is difficult to definitively state that there is no
extended low energy tails due to high background rate in instruments such as GBM or BATSE. 
\citet{2002ApJ...567.1028C} reported the existence of
extended low energy prompt emission tails in GRBs detected with BATSE. In this case
the background is averaged over several orbits and then subtracted from the data in order to improve the statistical significance.
Similar work, using the GBM data is currently in progress \citep{2011arXiv1111.3779F}.

Estimation of the number of GRBs that the LAT should detect based on
extrapolation of the high energy power law when fitting the keV--MeV emission
with a Band function are usually too optimistic compared to the real
observations. This suggests that the additional power law do not exist in all
GRBs, and that the Band function is a too simplistic model to describe the
prompt emission spectral shape. Whether a cut off is required around few tens of
MeV (Abdo et al., in preparation), 
which could be explained by emission process or pair
creation opacity, or as reported in \citet{2011ApJ...727..L33}, the addition of
an other component such as a black body to the Band function modify the
parameters of the latest making $\beta$ steeper and more compatible with the
number of GRB detected by the LAT. Moreover, the black body component reported
in \citet{2011ApJ...727..L33} also makes $\alpha$ steeper, and thus, more
compatible with synchrotron models.  The consistency of GBM data with synchrotron
spectra is also discussed in \citet{2011ApJ...741...24B}, who report the fit of
the prompt emission spectrum of GRB 090820A with an analytical synchrotron model,
and conclude on the good fit when when the data are simultaneously fit with a synchrotron function
and with a black body component. 
\citet{2009A&A...498..677B, 2011A&A...526A.110D} and \citet{2009ApJ...703..675N} 
investigated the effect of inverse-Compton scattering off the synchrotron spectrum on the 
resulting observed spectrum and conclude that the data could be reconciled with the synchrotron emission models.

\subsubsection{Delayed high energy emission}\label{sss:he_delay}

\begin{figure}
\centerline{\includegraphics[width=11.0cm]{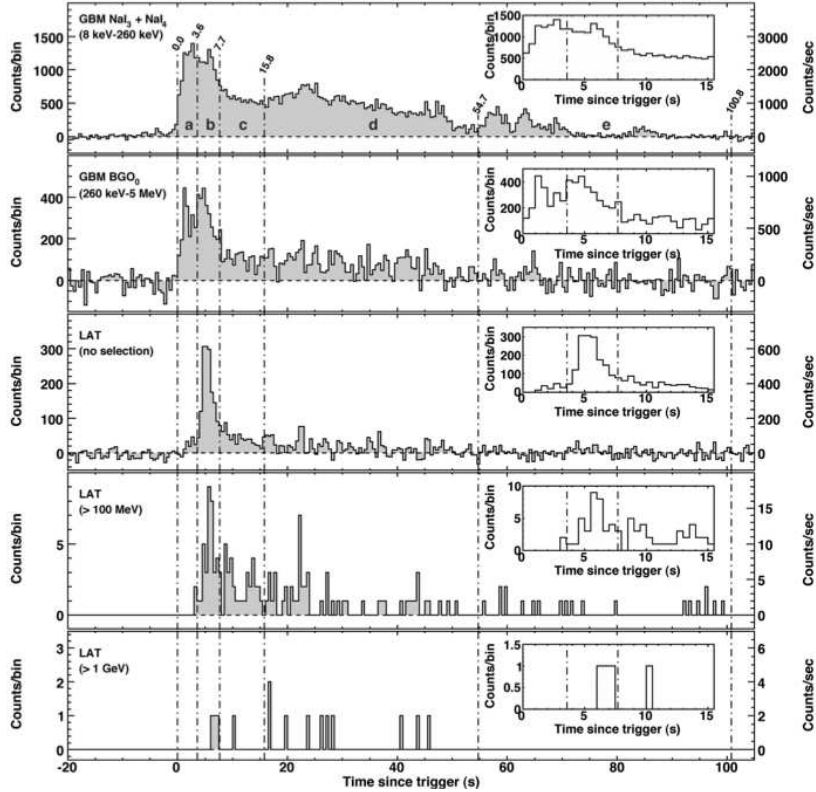}}
\caption{Energy dependent optical depth for $\gamma$-rays from the
GRBs.\label{f:grb1_lcs}}
\end{figure}

In the long, bright and hard GRB\,080825C triggered by GBM and seen by LAT,
high-energy $\gamma$ ray emission detected by the LAT starts later and persist
longer than the lower energy $\gamma$ ray photons. This is also seen
subsequently in the long GRB\,080916C, as well as short bursts, GRB\,081024B
and GRB\,090510C (see LAT catalog). Therefore they appear to be a common
feature of bright and hard GRBs seen by both GBM and  LAT. The delayed onset of
the GRB\,080916C LAT pulse, which coincides with the rise of the second peak in
the GBM light curve (see Fig.~\ref{f:grb1_lcs}), suggests a common origin in
a region spatially separated from the first GBM pulse. In the framework of the
internal-shock model for the prompt emission of GRBs where intermittent
relativistic shells of plasma are ejected by a newly formed black hole and
collide to form shocks and accelerate particles, the two emission regions could
arise from two different pairs of colliding shells, with variations in physical
conditions leading to non-thermal electrons with different spectral hardnesses
\citep{1999PR...314..575, 2002ARAA...40..137}. Model based on the proton
synchrotron radiation in the prompt phase where the delay is due to
acceleration of protons, also is consistent with the  high energy spectrum of
GRB\,080916C \citep{2010OAJ.....3..150R}.

\begin{figure}
\centerline{\includegraphics[width=11.0cm]{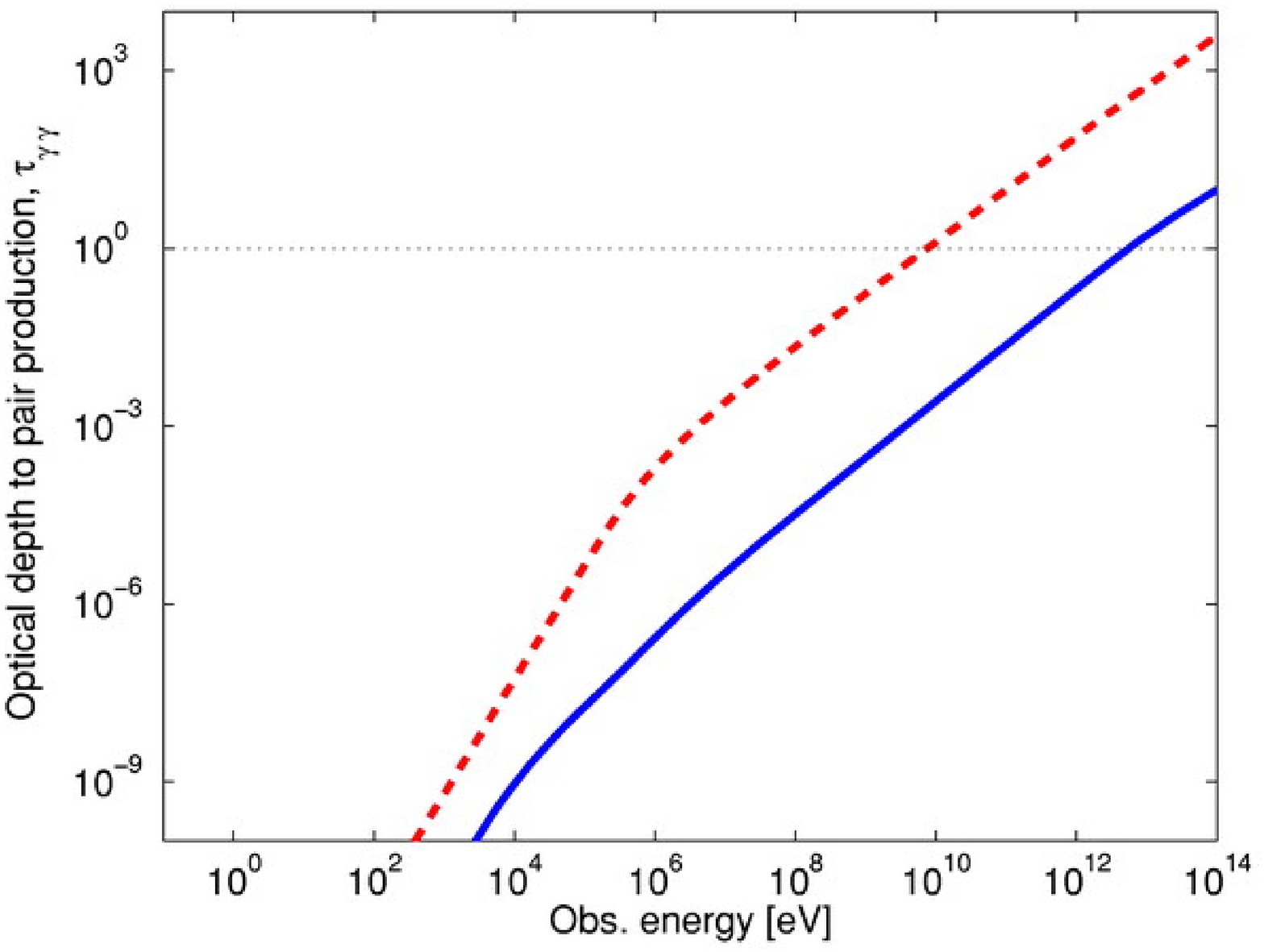}}
\caption{Energy-dependent optical depth to pair production, for the two
scenarios. Solid line, Explosion into ISM; dashed line, explosion into a
wind.\label{f:en_tau}}
\end{figure}

An alternative explanation for the delayed onset of the LAT emission is that a
volume becomes filled with radiation that attenuates the high-energy photons
until a later time when the emitting region expands and becomes optically thin
\citep{2008ApJ...677...92,arXiv:1107.5737}. \citet{2005IJMPA...20..3163} have predicted that
$\gamma$-ray photons typically above 10--100~GeV will be trapped inside the
fireball due to a high opacity of electron-positron pair production with other
photons. High energy photons escaping the fireball may interact with cosmic
background radiation and provide delayed $\gamma$-ray emission.
Fig.~\ref{f:en_tau} shows the photon energy dependence of the optical depth
inside the fireball \citep{2005ApJ...633..1018}. A $\gamma\gamma$
pair-production opacity effect would, however, produce a high-energy spectral
softening or cutoff, whereas in all cases the combined GBM/LAT data are well
fit with simple models by using the Band parameterisation.

For GRB\,090510 an electron synchrotron radiation model in the early afterglow
has been invoked \citep{2009MNRAS.400L..75K, 2010A&A...510L...7G}.

\section{Inferred results}\label{s:inf_res}

\subsection{Standard candle}\label{ss:std_candle}

During their rather short live time, GRBs are radiating more energy than all
the rest of the Universe and could therefore be the farthest objects, which
could be possible to detected if they can actually be produced by the first
generation of stars and if there is not too much absorption along the line of
sight. Similar to type Ia supernovae, if GRBs are standard-like candles, they
could supplement the former sample and could be the deepest probes of the
cosmos. Lots of efforts have been put into testing hardness-intensity relations
in GRB prompt emission. First proposed by \citet{2002A&A...390...81A} from the
BeppoSAX data, the so-called `Amati relation' relates the observed $E_{\rm
peak}$ to the total isotropic energy release. Hence this raised lots of
discussion in the community. This relation suggests a correlation between the
parameter $E_{\rm peak}$ of the Band function and the total energy radiated
during the prompt phase (when corrected for the redshift).
\[
  E_{\rm peak} \propto E_{\rm iso}^{(0.52\pm 0.06)}.
\]
The caution on the validity of this relation is the possible instrumental and
selection effects. Since the first proposition in 2002, many other derived
results have been proposed. Among them the so called Ghirlanda relation is
particularly interesting. It states that the collimation-corrected total energy
of the bursts, $E_\gamma$ is well correlated with the source frame $E_{\rm
peak}$ as follows:
\[
  E^{\rm obs}_{\rm peak}(1+z) \propto E_\gamma^{0.7}.
\]

The physics of GRBs is far from understood, these relations are purely
empirical but survive from one generation of instrument to the other
\citep{2003ChJAS...3..455A, 2006ApJ...636L..73S}. More recently,
\citet{2010APJ...721..1329} re-analyzed the full sample of BATSE GRBs, and
showed that most of the GRBs violate the limits imposed by the Amati relation,
but are consistent with the Ghirlanda relation \citep{2004ApJ...616..331}.
\citet{astro-ph:1110.6173} very recently showed through the use of a population
synthesis code to model the prompt gamma-ray emission from GRBs that a
combination of instrumental sensitivity and the cosmological nature of an
astrophysical population can artificially produce a strong correlation between
observed parameters like the Amati relation.

\subsection{Extra-galactic background light}

The Extra-galactic Background Light (EBL) is a cosmic diffuse Infra-red
(IR)-Optical-Ultra-violet (UV) radiation field produced by the first generation
of stars and its reprocessing by the dust in the interstellar medium. The EBL
is difficult to study directly because of the intense foreground and galactic
emission. However, it is possible to constrain indirectly the EBL by studying
its effect on high energy gamma ray emission. In fact, photons with energy
above $\sim 10$~GeV can interact with such a low energy photon field via
electron--positron pair production during their propagation from sources at
cosmological distances. This will result in a spectral break in the observed
spectrum from high energy sources like the Blazars and GRBs which are ideal
candidates to study EBL. Thus it is possible to test the density of the EBL
photon field by studying its opacity to high energy photons. Study of EBL as a
function of the redshift could lead to  information on galaxy evolution and
star formation in the early Universe. The highest energy photons observed with
Fermi, such as the 33.4 GeV photon from GRB\,090902B at a redshift of $z=1.82$,
do not support the `fast evolution' and the `baseline' models because they
predict optical depths of $\tau_{\gamma\gamma}=7.7$ and 5.8 respectively which
is too thick to support the observation \citep{2006ApJ...648..774S}. Similarly
EBL models predicting a greater opacity of the Universe to high energy
$\gamma$-rays, in the GeV--TeV, energy range are in disagreement with the
observations of Blazars \citep{2006ApJ...648..774S, 2007ApJ...658.1392S}.
Results on the measurement of the opacity of the Universe with Fermi are
reported in \citet{2009ApJ...706L.138A, 2010ApJ...723.1082A}.

\subsection{Lorentz invariance violation}

While special relativity assumes that there is no fundamental length-scale
associated with Lorentz invariance, some quantum gravity models predict a
fundamental length scale (called the Planck scale $l_{\rm Planck} \approx
1.62\times10^{-33}$~cm or $E_{\rm Planck} \approx M_{\rm Planck} c^2 \approx
1.22 \times 10^{19}$ GeV) at which quantum effects are expected to strongly
affect the nature of space-time. This implies that the Lorentz invariance may
break at or near the Planck scale. Some quantum gravity models predict a
possible test of such a violation is to measure an energy-dependence of the
speed of light. To detect such an effect, photons have to propagate very large
distances to make possible the measurement of any time delay between photons of
various energies emitted at the same time from the source. GRBs are explosive
events at cosmological distances, emitting photons over a very large energy
range. Although the emission mechanisms involved in GRBs are still not
completely understood, it is generally accepted that low and high energy
photons are emitted simultaneously from the same emission region and that no
known delay can be attributed to any source related emission effects, which is
a strong hypothesis that could impact the validity of the results. Any delay
between low and high energy photons in the observer frame could then be
attributed to photon propagation effects. Most of the models predicting a
Lorentz invariance violation conclude that the high energy photons should be
delayed with respect to low energy photons, but the opposite could happen as
well. The arrival time delay (i.e.\ lag), $\Delta t$, between low and high
energy photons, $E_{\rm l}$ and $E_{\rm h}$ respectively, emitted
simultaneously and at the same location, is given by \citep{1998Nat...393..763}
\[
  \Delta t =\xi \left( \frac{E_{\rm h}-E_{\rm l}}{M_{\rm Planck} c^2} \right)
     \frac{L}{c}
\]
(using only linear terms in the dispersion relation) where $\xi$ is a
model-dependent factor of the order of unity and $L$ the distance to the
source.

All quantum gravity models do not agree on the degree of possible violation.
With Fermi, due to its unprecedented energy bandwidth, first order and possibly
second order effect could be measured. \citet{2009Sci...323.1688} using
GRB\,090510 determined lower limits on the quantum gravity mass scale. The
technique described previously was applied assuming standard cosmology of a
flat expanding Universe with the following parameters $\Omega_{M}=0.27$,
$\Omega_{\Lambda}=0.73$ and $H_0=71$ km~s$^{-1}$Mpc$^{-1}$. Very conservative
constraints were provided considering the lag as the duration between the onset
of the burst at low energy and the arrival time of the highest energy photons.
In the case of the long GRB\,080916C, the highest energy photon is a 13 GeV
arriving 16.5~s after the GRB onset and for the short GRB\,090510, a 31 GeV
photon was observed by the LAT 0.83~s after the GBM trigger time. The strongest
constraint was obtained for GRB\,090510 with $E_{\rm LIV} \geq 1.19 E_{\rm
Planck}$. Even stronger constraints were obtained using less conservative but
still reasonable assumptions, by associating after-trigger low energy pulses
with the high energy photons.
%

Until now, no significant energy dependence of the speed of light was reported
from the Fermi data, which currently provides the best lower limits. This
strongly disfavours the quantum gravity models predicting that the granularity
of the space-time at Planck scale should lead to a linear dependence of the
speed of light with the energy.

\section{Future prospects}\label{s:future}

\subsection{VHE and UHE Gamma-rays from GRB Sources}\label{ss:vhe}

The terrestrial atmosphere provides nearly 28 radiation lengths of shielding to
celestial $\gamma$-rays. Ground based telescopes like the Imaging Atmospheric
\v{C}erenkov Telescopes (IACTs) for very high energy (VHE) $\gamma$-ray
astronomy and the Extensive Air Shower (EAS) arrays for PeV $\gamma$-ray
astronomy use the atmosphere itself as the detector of $\gamma$-rays. These
experiments exploit the characteristics of the secondaries of the
electromagnetic cascade initiated by the $\gamma$-rays.

IACTs such as MAGIC, VERITAS and HESS, which are still operational and water
\v{C}erenkov detectors such as MILAGRO, whose activity stopped in 2008, looked
for and are still looking for a possible VHE $\gamma$ ray signal in the GeV/TeV
regime coming from GRBs. More recently MAGIC \citep{2010AIPC.1279..312G},
VERITAS \citep{2009arXiv0907.4997G} and HESS \citep{2009AAP...495..505A} come
online. Until now, only upper limits have been reported for a few tens of GRBs.
MAGIC, with a low energy threshold ($\sim 25$~GeV) and the fastest repointing
capabilities is currently the most powerful of these instruments for a quick
follow up of the VHE counterpart in GRBs in response to alerts from other
instruments provided through GCN. However, MAGIC cannot repoint faster
than a few tens of seconds, which is usually too long to catch the prompt
emission from the burst. While IACTs suffer from a handicap of low duty cycles
and limited field of views, water \v{C}erenkov detectors or the EAS arrays have
a full sky coverage and can operate during day and night with no limitation
with the Moon's phase. MILAGRITO, a prototype version of MILAGRO, reported a
possible VHE $\gamma$ ray signal in coincidence with GRB\,970417A, which
triggered BATSE \citep{2000ApJ...533L.119A}. However, MILAGRO, a much more
sensitive instrument, unfortunately did not find such a detection
\citep{2005ApJ...630..996A}.

Even though IACTs and water \v{C}erenkov detectors have not succeeded in
detecting a VHE $\gamma$-ray signal from the GRBs they were only a small number
of GRBs. MAGIC energy threshold overlaps with the LAT energy band and hence
could have detected the 33 GeV photon observed by the LAT during the prompt
emission of GRB090902B \citep{2009ApJ...706L.138A}. In addition, this photon
was detected about 80~s after the GBM trigger time, which is encouraging since
MAGIC could have repointed to the burst location during this time. If this
burst were to have triggered Swift or LAT and provided a precise location
through the GCN system the chances could have been still higher. Future
generation IACTs, like the \v{C}erenkov Telescope Array (CTA)
\citep{2011arXiv1109.5680B}, and the future water \v{C}erenkov detector like
the High-Altitude Water \v{C}erenkov Observatory 
(HAWC), if operating simultaneously with Fermi, should be able to undertake the
challenge of detecting VHE $\gamma$-ray emission from GRBs. Such a detection
will have a major impact on GRB physics. The high energy cutoff, if any, could
tell us about the emission process operating at the GRB sources, the
acceleration mechanism of the radiating particles, the outflow Lorentz factor
as well as the Extra-galactic Background Light (EBL) responsible for VHE
extinction \citep{2010ApJ...723.1082A}.

\subsection{Radio emission from GRB sources}\label{ss:radio}

A conducting fireball expanding at relativistic speed into an ambient magnetic
field generates a rapidly changing electric current which emits coherent
electromagnetic radiation at radio frequencies. The critical frequency (upper
limit of the emission) strongly depends on the Lorentz factor of the expansion.
The radio detection will enable us to estimate the density and structure of the
circumstellar material and, by inference, the evolution of the presupernova
stellar wind, and reveal the last stages of stellar evolution before explosion.
The radio emission can best be explained as the interaction of a mildly
relativistic ($\Gamma \sim 1.6$) shock with a dense pre-explosion stellar
wind-established circumstellar medium that is highly structured both
azimuthally, in clumps or filaments, and radially, with observed density
enhancements.

Over the past several years the afterglow of $\gamma$-ray bursters has
occasionally been detected in the radio
(see Chandra \& Frail 2011 (this issue) for a comprehensive review on radio 
afterglow observations from GRBs), as well in other wavelengths.  It is
possible to model the gross properties of the radio and optical/infrared
emission from the half-dozen GRBs with extensive radio observations. From this
it is concluded that at least some members of the `slow-soft' class of GRBs can
be attributed to the explosion of a massive star in a dense, highly structured
circumstellar medium that was presumably established by the pre-explosion
stellar system.

The coherent radio emission is practically simultaneous with the GRB except for
the reduced propagation speed of the radio waves by interstellar dispersion.
However delayed radio emissions have been detected in the case of GRB\,990123
\citep{1999ApJ...522..L97} which has been interpreted as due to reverse shock.
The source of this radio emission could be attributed to incoherent synchrotron
radiation of shockwave produced electrons \citep{1993ApJ...418..L5}.

Ideally an all sky monitor in the radio frequency range would be an ideal
instrument since unknown delays due to dispersion measure as well as intrinsic
delays due to the emission process. One such possible detector would be the
upcoming LOw Frequency ARray (LOFAR) array which has a sensitivity at low
frequencies ($\leq~250$ MHz) better than the currently operating radio
telescopes like the Very Large Array (VLA), Westerbork Synthesis Radio
Telescope (WSRT) and Giant Meterwave Radio Telescope (GMRT) array, by more than
an order of magnitude.

In addition the Square Kilometer Array (SKA) will be a revolutionary radio
telescope with about one square kilometre of collecting area, giving 50 times
the sensitivity and 10,000 times the survey speed of the best current day
telescopes with a wide bandwidth of frequency range from 70 MHz to 10 GHz. It
is expected to become fully operational by 2024. We would certainly expect
breakthroughs when these future projects become fully operational.

\subsection{Neutrinos from GRBs}\label{ss:neutr}

Detection of high energy neutrinos from GRBs would test the shock acceleration
mechanism and the suggestion that GRBs are the sources of ultra-high energy
protons, since $\geq 10^{14}$ eV neutrino production requires $10^{16}$ eV
protons. The dependence of even higher energy neutrino ($10^{17}$ eV) flux on
the fireball environment implies that the detection of high energy neutrinos
would also provide constraints on the GRB progenitors. Furthermore, it has
recently been pointed out that if GRBs originate from core-collapse of massive
stars, then a burst of $\geq 5$ TeV neutrinos may be produced by photo-meson
interaction while the jet propagates through the envelope with TeV fluence
implying 0.1--10 neutrino events per individual collapse in a 1~km$^3$ neutrino
telescope. It is argued that GRBs have the potential to produce the particle
energies (up to $10^{21}$ eV) and the energy budget ($10^{44}$
erg~yr$^{-1}$Mpc$^{-3}$) to accommodate the spectrum of the highest energy
cosmic rays (see \S\ref{ss:uhecr} for more details). Detection of neutrinos is
the only unambiguous way to establish that GRBs accelerate protons.

Detection of neutrinos from GRBs could be used to test the simultaneity of
neutrino and photon arrival to an accuracy of $\sim 1$~s checking the
assumption of special relativity that photons and neutrinos have the same
limiting speed, especially in view of the recent results from the OPERA
experiment at Gran Sasso \citep{arXiv.1109.4897}.

Hence it is important to search for neutrino emission from GRBs. Is the
predicted flux detectable by the existing neutrino telescopes?
\citet{1997PRL...78..2292} show that a large fraction, $\geq 10$\% of the
fireball energy is expected to be converted by photo-meson production to a
burst of $\sim 10^{14}$~eV neutrinos. As a result we would expect to see
several tens of events in a year in a typical neutrino detector of area
2~km$^2$. This rate is comparable to the background expected from atmospheric
neutrinos. However one could easily detect a burst of neutrinos temporally and
directionally coincident with the $\gamma$-rays from the GRB. The predicted
neutrino flux also implies a detection of $\sim 10$ neutrino induced muon
events per year in a planned 1~km$^3$ \v{C}erenkov neutrino detectors,
correlated in time and direction with GRBs. There are several systematic
searches for neutrino emission strongly time correlated with GRB signals
\citep{2000PhRvD...62..093015} produced in a Waxman--Bachall fireball
\citep{1997PRL...78..2292} as well as model independent searches for neutrinos
in a range of energies and emission times with no apparent success
\citep{2011AIPC...1358..351, 2011AIPC...1358..361, 2011ApJ...736...50}. Perhaps
this lack of  success may be understandable if the neutrinos travel with
super-luminal velocities \citep{arXiv.1109.5378}.

The future neutrino detector has the potential to answer some of these critical
questions.

\subsection{Gravitational waves from GRBs}\label{ss:gravity}

As predicted in Einstein's General Theory of Relativity, gravitational waves
are disturbances in the curvature of spacetime caused by the motions of matter.
A promising source of gravitational waves is the coalescence of compact binary
systems, events which are now believed to be the origin of short GRBs because
such mergers are asymmetric. In addition, short GRBs are generally closer than
long GRBs. Further GRBs provide a convenient time stamp to identify coincident
signals from the gravitational wave detectors which are not directional.
Sensitivities of some of the upcoming gravitational wave detectors mentioned
below are sufficient to allow some possibility for the detection of
astrophysical sources, including gravitational wave signals associated with
GRBs and SGRs. Propagating at (or near) the speed of light, gravitational waves
pass straight through matter, their strength weakens proportionally to the
distance travelled from the source. A gravitational wave arriving on Earth will
alternately stretch and shrink distances, though on an incredibly small scale
-- by a factor of $10^{-21}$ for very strong sources. Gravitational wave
astronomy could expand our knowledge of the cosmos dramatically. For example,
gravitational waves, though weakening with distance, are thought to be
unchanged by any material they pass through and, therefore, should carry
signals unaltered across the vast reaches of space. By comparison,
electromagnetic radiation tends to be modified by intervening matter. The
remarkable thing about a black hole when simulated on a computer is that no
matter how it forms or is perturbed, whether by infalling matter, by
gravitational waves, or via a collision with another object (including a second
black hole), it will `ring' with a unique frequency known as its natural mode
of vibration. It's this unique wave signature that will allow scientists to
know if they've really detected a black hole. The signal will tell how big the
black hole is and how fast it's spinning. The $\gamma$-rays and the afterglow
emission of GRBs are thought to be produced at distances from the central
engine where the plasma has become optically thin, $r \geq 10^{13}$ cm, which
is much larger than the Schwarzschild radius of a stellar mass black hole (or
of a neutron star). Hence we have only very indirect information about the
inner parts of the central engine where the energy is generated. However, in
any stellar progenitor model of GRB one expects that gravitational waves should
be emitted from the immediate neighbourhood of the central engine which is known
to house a black hole. Gravitational waves  emitted from the progenitor itself,
would carry more direct information on the properties of the central engine.
Therefore, it is of interest to study the gravitational wave emission from GRB
associated with specific progenitors.

Existence of a shallow decay phase in the early X-ray afterglows of
$\gamma$-ray bursts is a common feature. It is possible that this is connected
to the formation of a highly magnetised millisecond pulsar, pumping energy into
the fireball on timescales longer than the prompt emission. In this scenario,
the nascent neutron star could undergo a secular bar-mode instability, leading
to gravitational wave losses which would affect the neutron star spin-down. In
this case, nearby $\gamma$-ray bursts with isotropic energies of the order of
$10^{50}$ erg would produce a detectable gravitational wave signal emitted in
association with an observed X-ray light-curve plateau, over relatively long
timescales of minutes to about an hour. The peak amplitude of the gravitational
wave signal would be delayed with respect to the $\gamma$-ray burst trigger,
offering gravitational wave interferometers such as the advanced LIGO and Virgo
the challenging possibility of catching its signature on the fly
\citep{2009ApJ...702..1171}.

Gravitational waves if detected from well-localised, spiralling in
compact-object binaries, like the progenitors of short GRBs, can measure
absolute source distances with high accuracy. When coupled with an independent
determination of redshift through an electromagnetic counterpart, these
standard sirens can provide an excellent probe of the expansion history of the
Universe and eventually the dark energy equation of state parameter $w$
\citep{2006PhRvD..74f3006D}.

Hence the ground based gravitation wave detector network consisting of Laser
Interferometer Gravitational-Wave Observatory (LIGO), LIGO-II (a planned
upgrade of LIGO with tenfold increase in sensitivity), Virgo (Italian and
French collaboration for the realization of an interferometric gravitational
wave detector which is about to be fully commissioned), and AIGO are vital to
realize some the potentials of this technique. LIGO--VIRGO network will become
operational in 2015 at the completion of the latest upgrades. Prime candidates
for generating detectable radiation are binary black holes whose constituents
each have masses of 10--50~$M_\odot$.
%
%
In addition there are several gravitational wave detector arrays planned for
the future. Australian Consortium for Interferometric Gravitational wave
Astronomy (ACIGA), plans at undertaking research and development aimed at
improving the performance of present laser interferometric gravitational wave
detectors through advanced designs to ultimate limits set by mechanics, quantum
mechanics, lasers and optics. The Einstein Telescope (ET) aims at realization
of the conceptual design of a future European third generation gravitational
wave detector. In fact the evolution of the current (first generation)
gravitational wave detectors is well defined: after the current upgrade to the
so-called enhanced level, the detectors will evolve toward their second
generation: the advanced Virgo and LIGO detectors.

There are several searches conducted already for gravitational wave signal
coincident with GRBs with no clear success \citep{2008ApJ...681.1419,
2008CQGra..25v5001, 2008AIPC...1000..284}. With the commissioning of more
advanced detectors there is potential for achieving breakthroughs in the field
of Gravitational Wave Astronomy (see Dhurandhar 2011).

\section*{Acknowledgements}

SG was supported by an appointment to the NASA Postdoctoral Program at the
Goddard Space Flight Center, administered by Oak Ridge Associated Universities
through a contract with NASA.


\label{lastpage}

\begin{thebibliography}{}
%
\bibitem[Abbott et~al.(2008)]{2008ApJ...681.1419}
  Abbott B., et~al., 2008, ApJ, 681, 1419
%
\bibitem[Abdo et~al.(2009a)]{2009Sci...323.1688}
  Abdo A.~A., et~al., 2009a, Science, 323, 1688
%
%
\bibitem[Abdo et~al.(2009c)]{2009ApJ...706L.138A}
  Abdo A.~A., et~al., 2009c, ApJ, 706, L138
%
\bibitem[Abdo et~al.(2009b)]{2009ApJ...707..580}
  Abdo A.~A., et~al., 2009b, ApJ, 707, 580
%
\bibitem[Abdo et al.(2010a)]{2010ApJ...712..558A}
  Abdo A.~A., et al., 2010a, ApJ, 712, 558
%
\bibitem[Abdo et~al.(2010b)]{2010ApJ...723.1082A}
  Abdo A.~A., et~al., 2010b, ApJ, 723, 1082
%
%
\bibitem[Acernese et~al.(2008)]{2008CQGra..25v5001}
  Acernese F., et~al., 2008, Classical and Quantum Gravity, 25, 225001
%
\bibitem[Ackermann et~al.(2011)]{2011ApJ...729..114A}
  Ackermann M., et~al., 2011, ApJ, 729, 114
%
\bibitem[Ackermann et~al.(2010)]{2010ApJ...716.1178A}
  Ackermann M., et~al., 2010, ApJ, 716, 1178
%
\bibitem[Adam et~al.(2011)]{arXiv.1109.4897}
  Adam T., et~al., 2011, arXiv:1109.4897 [hep-ph]
%
\bibitem[Aharonian et~al.(2009)]{2009AAP...495..505A}
  Aharonian F., et~al., 2009, A\&A, 495, 505
%
\bibitem[Alexander(2008)]{2008AIPC...1000..284}
  Alexander D., 2008, in Gamma-ray bursts 2007, eds, Galassi M., Palmer D., Fenimore E., AIP Conf. Proc., 1000, 284
%
\bibitem[Alvarez-Muniz, Halzen \& Hooper(2000)]{2000PhRvD...62..093015}
  Alvarez-Muniz J., Halzen F., Hooper D. W., 2000, Phys. Rev. D62, 093015
%
\bibitem[Amati(2003)]{2003ChJAS...3..455A}
  Amati L.,
    2003, Chinese Journal of Astronomy and Astrophysics Supplement, 3, 455
%
\bibitem[Amati et~al.(2002)]{2002A&A...390...81A}
  Amati L., et~al., 2002, A\&A, 390, 81
%
\bibitem[Amelino-Camelia et~al.(1998)]{1998Nat...393..763}
  Amelino-Camelia G., Ellis J., Mavromatos N.E., Nanopoulos D. V., Sarkar S.,
    1998, Nature, 393, 763
%
\bibitem[Aptekar et~al.(1995)]{1995SSR...71..265}
  Aptekar R. L., et~al.,
    1995, Space Science Reviews, 71, 265
%
\bibitem[Asano, Guiriec \& M{\'e}sz{\'a}ros(2009)]{2009ApJ...705L.191A}
  Asano K., Guiriec S., M{\'e}sz{\'a}ros P.,
    2009, ApJ, 705, L191
%
\bibitem[Atkins et~al.(2000)]{2000ApJ...533L.119A}
  Atkins R., et~al., 2000, ApJ, 533, L119
%
\bibitem[Atkins et~al.(2005)]{2005ApJ...630..996A}
  Atkins R., et~al., 2005, ApJ, 630, 996
%
\bibitem[Atwood et~al.(2009)]{2009ApJ...697..1071}
  Atwood W. B., et al., 2009, ApJ, 697, 1071
%
\bibitem[Autiero, Migliozzi \& Russo(2011)]{arXiv.1109.5378}
  Autiero D., Migliozzi P., Russo A.,
    2011, arXiv:1109.5378 [hep-ph]
%
%
\bibitem[Band et~al.(1993)]{1993ApJ...413..281B}
  Band D., et~al., 1993, ApJ, 413, 281
%
\bibitem[Bednarz \& Ostrowski(1998)]{1998PRL...80..3911}
  Bednarz J., Ostrowski M., 1998, Phys. Rev. Lett., 80, 3911
%
\bibitem[Beloborodov(2010)]{2010MNRAS.407.1033B}
 Beloborodov, A.~M.,
  2010, MNRAS, 407, 1033
%
\bibitem[Bhat et~al.(1992)]{1992NAT...359..217}
  Bhat P.~N., Fishman G. J., Meegan C. A., Wilson R. B., Brock M. N., Paciesas W. S.,
    1992, Nature, 359, 217.
%
\bibitem[Bhat et~al.(1994)]{1994ApJ...426..604}
  Bhat P.~N., Fishman G. J., Meegan C. A., Wilson R. B., Kouveliotou C., Paciesas W. S., Pendleton G. N.,
  Schaefer B. E., 1994, ApJ, 426, 604
%
%
\bibitem[Bhat et~al.(2011)]{2011ApJ...741}
  Bhat P.~N., et~al., 2011, ApJ, 741, in press, arXiv:1109.4064 [astro-ph]
%
\bibitem[Bissaldi et~al.(2011)]{2011ApJ...733...97}
  Bissaldi E., et al., 2011, ApJ, 733, 97
%
\bibitem[Blaufuss, Meagher \& Whitehorn(2011)]{2011AIPC...1358..351}
  Blaufuss E., Meagher K., Whitehorn N., 2011, in Gamma Ray Bursts 2010, eds McEnery J.E.,
   Racusin J.L., Gehrels N., AIP Conf. Proc., 1358, 351
%
\bibitem[Bloom et~al.(2006)]{2006ApJ...638..354}
  Bloom J. S., et~al., 2006, ApJ, 638, 354
%
\bibitem[Bloom et~al.(1998)]{1998ApJ...507..L25}
  Bloom J. S., Djorgovski S. G., Kulkarni S. R., Frail D. A.,
    1998, ApJ, 507, L25
%
\bibitem[Bonnell et~al.(1997)]{1997ApJ...490...79}
  Bonnell J. T., Norris J. P., Nemiroff R. J., Scargle J. D.,
    1997, ApJ, 490, 79
%
\bibitem[Bo{\v s}njak et al.(2009)]{2009A&A...498..677B} 
Bo{\v s}njak, {\v Z}., Daigne, F., \& Dubus, G.\ 
2009, A\&A, 498, 677 
%
\bibitem[Bouvier(2010)]{2010Astro...1012..0558}
  Bouvier A.,
    2010, Ph.D. Thesis, arXiv:1012.0558 [astro-ph]
%
\bibitem[Bouvier et~al.(2011)]{2011arXiv1109.5680B}
  Bouvier A., Gilmore R., Connaughton V., Otte N., Primack J.R., Williams D.A.,
    2011, arXiv:1109.5680 [astro-ph.HE]
%
\bibitem[Briggs et~al.(1999)]{1999ApJ...524..82B}
  Briggs M.~S., et~al., 1999, ApJ, 524, 82
%
\bibitem[Burgess et al.(2011)]{2011ApJ...741...24B}
 Burgess, J.~M., Preece, R.~D., Baring, M.~G., et~al., 2011, ApJ, 741, 24 
%
\bibitem[Cabrera et~al.(2007)]{2007MNRAS.382..342}
  Cabrera J. I., Firmani C., Avila-Reese V., Ghirlanda G., Ghisellini G.,
  Nava L.,
    2007, MNRAS, 382, 342
%
\bibitem[Cavallo \& Rees(1978)]{1978MNRAS...183..359}
  Cavallo G., Rees M.~J.,
    1978, MNRAS, 183, 359
%
\bibitem[Chandra \& Frail(2011)]{2011BASI...39..143}
Chandra P., Frail D.A., 2011, BASI, 39, in press
%
\bibitem[Chen et~al.(2005)]{2005ApJ...619..983}
  Chen L., Lou Y-Q., Wu M., Qu J-L., Jia S-M., Yang X-J.,
    2005, ApJ, 619, 983
%
\bibitem[Cline et~al.(2003)]{2003AIP...662..143}
  Cline T. L., et~al., 2003, in Gamma-Ray Burst and Afterglow Astronomy 2001, eds
  Ricker G.R., Vanderspek R.K., AIP Conf. Proc., 662, 143
%
\bibitem[Connaughton(2002)]{2002ApJ...567.1028C}
 Connaughton, V.,
  2002, ApJ, 567, 1028 
%
\bibitem[Corsi \& M{\'e}sz{\'a}ros(2009)]{2009ApJ...702..1171}
  Corsi A., M{\'e}sz{\'a}ros P.,
    2009, ApJ, 702, 1171
%
\bibitem[Costa et~al.(1997)]{1997Nat...387..783}
  Costa E., et~al., 1997, Nature, 387, 783
%
\bibitem[Crider et~al.(1997)]{1997ApJ...479L..39C}
  Crider A., et~al.,
    1997, ApJ, 479, L39
%
\bibitem[Cucchiara, Levan \& Fox(2011)]{2011ApJ...736....7}
  Cucchiara A., Levan A. J., Fox D. B., 2011, ApJ, 736, 7
%
\bibitem[Daigne \& Mochkovitch(1998)]{1998MNRAS...296..275}
  Daigne F., Mochkovitch R.,
    1998, MNRAS, 296, 275
%
\bibitem[Daigne \& Mochkovitch(2002)]{2002MNRAS.336.1271D}
  Daigne F., Mochkovitch R., 2002, MNRAS, 336, 1271
%
\bibitem[Daigne \& Mochkovitch(2003)]{2003MNRAS...587..592}
  Daigne F., Mochkovitch R.,
    2003, MNRAS, 587, 592.
%
\bibitem[Daigne et al.(2011)]{2011A&A...526A.110D}
 Daigne, F., Bo{\v s}njak, {\v Z}., \& Dubus, G.\ 
 2011, A\&A, 526, A110

%
\bibitem[Dalal et~al.(2006)]{2006PhRvD..74f3006D}
  Dalal N., Holz D. E., Hughes S. A., Jain B.,
    2006, Phys. Rev. D, 74, id. 063006
%
\bibitem[Davidson, Bhat \& Li(2011)]{2011AIPC...1358..17}
  Davidson R., Bhat P. N., Li G.,
    2011, in Gamma Ray Bursts 2010, eds McEnery J.E.,
   Racusin J.L., Gehrels N., AIP Conf. Proc., 1358, 17
%
\bibitem[Dermer(1998)]{1998ApJ...501..L157}
  Dermer C. D.,
    1998, ApJ, 501, L157
%
\bibitem[Dermer(2002)]{2002ApJ...574..65}
  Dermer C. D.,
    2002, ApJ, 574, 65
%
\bibitem[Dermer(2004)]{2004ApJ...614..284}
  Dermer C. D.,
    2004, ApJ, 614, 284
%
\bibitem[Dermer \& Mitman(1999)]{1999ApJ...513..L5}
  Dermer C. D., Mitman K. E.,
    1999, ApJ, 513, L5
%
%
   \bibitem[Dhurandhar(2011)]{2011BASI...39..181}
   Dhurandhar S.V., 2011, BASI, 39, 181
%
%
\bibitem[Djorgovski et~al.(2003)]{2003SPIE.4834..238}
  Djorgovski S. G., et~al.,
   2003, Proc. of the SPIE, 4834, 238
%
\bibitem[Fenimore, Madras \& Nayakshin(1996)]{1996ApJ...473..998}
  Fenimore E., Madras C. D., Nayakshin, S.,
    1996, ApJ, 473, 998
%
\bibitem[Fishman \& Meegan(1995)]{1995ARAA...33..415}
  Fishman G.~J., Meegan C.~A.,
    1995, ARA\&A, 33, 415
%
\bibitem[Fishman et~al.(1994)]{1994APJS...92..229}
  Fishman G.~J., et~al.,
    1994, ApJS, 92, 229
%
\bibitem[Fitzpatrick et al.(2011)]{2011arXiv1111.3779F} Fitzpatrick, G., Connaughton, V., McBreen, S., \& Tierney, D., 2011, arXiv:1111.3779
%
\bibitem[Foley et~al.(2011)]{2011AIPC...1358..183}
  Foley S., Bhat P. N., Gruber D., McBreen S., Tierney D., Greiner J., 
    2011, in Gamma Ray Bursts 2010, eds McEnery J.E.,
   Racusin J.L., Gehrels N., AIP Conf. Proc., 1358, 183
%
\bibitem[Ford et~al.(1995)]{1995ApJ...439..307F}
  Ford L.~A., et~al.,
    1995, ApJ, 439, 307
%
\bibitem[Frail et~al.(2001)]{2001ApJ...562..L55}
  Frail D. A., et~al.,
    2001, ApJ, 562, L55
%
\bibitem[Frail et~al.(2003)]{2003AJ...125..2299}
  Frail D. A., Kulkarni S. R., Berger E., Wieringa M. H.,
    2003, AJ, 125, 2299
%
\bibitem[Frail et~al.(2004)]{2003Aph...0308189}
  Frail D. A., Metzger B. D., Berger E., Kulkarni S.R., Yost S.A., 
      2004, ApJ, 600, 828
%
\bibitem[Galante et~al.(2009)]{2009arXiv0907.4997G}
  Galante N., for the VERITAS Collaboration,
    2009, arXiv:0907.4997 [astro-ph.HE]
%
\bibitem[Garczarczyk et~al.(2010)]{2010AIPC.1279..312G}
  Garczarczyk M., et~al.,
   2010, in Deciphering the Ancient Universe with Gamma-Ray Bursts,
   eds, Kawai N., Nagataki S.,  AIP Conf. Proc., 1279, 312
%
\bibitem[Gehrels(1997)]{1997NCimB...112..11G}
  Gehrels N.,
    1997, Nuovo Cimento B Serie, 112, 11
%
\bibitem[Gehrels et~al.(2006)]{2006Nat...444..1044}
  Gehrels N., et~al.,
    2006, Nature, 444, 1044
%
\bibitem[Ghirlanda, Celotti \& Ghisellini(2003)]{2003AAP...406..879}
  Ghirlanda G., Celotti A., Ghisellini G.,
    2003, A\&A, 406, 879
%
\bibitem[Ghirlanda, Ghisellini \& Lazzati(2004)]{2004ApJ...616..331}
  Ghirlanda G., Ghisellini G., Lazzati D.,
   2004, ApJ, 2004, 616, 331
%
%
\bibitem[Ghirlanda, Ghisellini \& Nava(2010)]{2010A&A...510L...7G}
  Ghirlanda G., Ghisellini G., Nava L.,
    2010, A\&A, 510, L7
%
\bibitem[Goldstein et al.(2010)]{2010APJ...721..1329}
  Goldstein A., Preece R.D., Briggs M. S., 2010,
    ApJ, 721, 1329
%
\bibitem[Goldstein et~al.(2011)]{2011APJSG}
  Goldstein A., et~al.,
    2011, ApJS, submitted
%
\bibitem[Gonz{\'a}lez et~al.(2003)]{2003Nat...424..749}
  Gonz{\'a}lez M.~M., Dingus B.~L., Kaneko Y., Preece R.~D., Dermer C.~D., Briggs M.~S., 
    2003, Nature, 424, 749
%
\bibitem[Granot, Cohen-Tanugi \& do Couto e Silva(2008)]{2008ApJ...677...92}
  Granot J., Cohen-Tanugi J., do Couto e Silva E.,
    2008, ApJ, 677, 92
%
%
\bibitem[Gruber et~al.(2011)]{2011AAP...531..20}
  Gruber D., et~al.,
    2011, A\&A, 531, 20
%
\bibitem[Guiriec et~al.(2010)]{2010ApJ...725..225G}
  Guiriec S., et~al.,
    2010, ApJ, 725, 225
%
\bibitem[Guiriec et~al.(2011)]{2011ApJ...727..L33}
  Guiriec S., et~al.,
   2011, ApJ, 727, L33
\bibitem[Guiriec et al.(2011)]{2011AIPC.1358...33G} Guiriec, S., 
  Connaughton, V., Briggs, M., et al.\ 2011, American Institute of Physics Conference Series, 1358, 33
%
\bibitem[Gupta, Das Gupta \& Bhat(2000)]{2000AIPC...526..215}
  Gupta V., Das Gupta P., Bhat P. N.,
    2000, in Gamma-ray bursts: 5th Huntsville symposium, eds, Kippen R.M., Fishman G.J., MallozziR.S.,
    AIP Conf. Proc., 526, 215
%
\bibitem[Gupta, Das Gupta \& Bhat(2002)]{2002Astro-ph...0206402}
  Gupta V., Das Gupta P., Bhat P. N.,
    2002, arXiv:astro-ph/0206402
%
\bibitem[Hakkila \& Preece(2011)]{2011ApJ...740..104}
  Hakkila J., Preece R. D.,
    2011, ApJ, 740, 104
%
\bibitem[Hakkila et~al.(2008)]{2008ApJ...677..L81}
  Hakkila J., Giblin T.W., Norris J. P., Fragile P. C., Bonnell J. T.,
    2008, ApJ, 677, L81
%
\bibitem[Hakkila \& Cumbee(2009)]{2009AIPC...1133..379}
  Hakkila J., Cumbee R. S.,
    2009, in Gamma-ray burst: Sixth Huntsville symposium, eds, 
    Meegan C., Kouveliotou C., GehrelsN.,  AIP Conf. Proc., 1133, 379
%
\bibitem[Hasco{\"e}t et~al.(2011)]{arXiv:1107.5737}
Hasco{\"e}t, R., Daigne F., Mochkovitch R., Vennin V.,
 2011, MNRAS, submitted , arXiv:1107.5737
%
\bibitem[Horv{\'{a}}th(2002)]{2002A&A...392..791H}
  Horv{\'{a}}th I.,
    2002, A\&A, 392, 791.
%
\bibitem[Hurley et~al.(1994)]{1994Nat...372..652}
  Hurley K., et~al.,
    1994, Nature, 372, 652
%
\bibitem[Ioka \& Nakamura(2001)]{2001ApJ...554..L163}
  Ioka K., Nakamura T.,
    2001, ApJ, 554, L163
%
\bibitem[Johnson et~al.(1993)]{1993ApJS...86..693}
  Johnson W. N., et~al.,
    1993, ApJS, 86, 693
%
\bibitem[Katz(1997)]{1997ApJ...490..663}
  Katz J. I.,
    1997, ApJ, 490, 663
%
\bibitem[Kirk et~al.(2000)]{2000ApJ...542..235}
  Kirk J. G., et~al.,
    2000, ApJ, 542, 235
%
\bibitem[Kobayashi, Piran \& Sari(1997)]{1997ApJ...490..92}
  Kobayashi S., Piran T., Sari R.,
    1997, ApJ, 490, 92
%
\bibitem[Kocevski \& Liang(2003)]{2003ApJ...594..385}
  Kocevski D., Liang E.,
    2003, ApJ, 594, 385
%
\bibitem[Kocevski(2011)]{astro-ph:1110.6173}
  Kocevski D.,
    2011, arXiv:1110.6173 [astro-ph.HE]
%
\bibitem[Kouveliotou et~al.(1993)]{1993ApJ...413L.101K}
  Kouveliotou C., Meegan C.~A., Fishman G.~J., Bhat P.N., Briggs M. S., Koshut T. M., Paciesas W. S., Pendleton G. N.,
    1993, ApJ, 413, L101
%
\bibitem[Kulkarni et~al.(1999)]{1999ApJ...522..L97}
  Kulkarni S. R., et~al.,
    1999, ApJ, 522, L97
%
\bibitem[Kumar \& Barniol Duran(2009)]{2009MNRAS.400L..75K}
  Kumar P., Barniol Duran R.,
    2009, MNRAS, 400, L75
%
\bibitem[Lazzati \& Begelman(2010)]{2010ApJ...725.1137L}
 Lazzati, D., \& Begelman, M.~C.,
 2010, ApJ, 725, 1137
%
\bibitem[Lemoine \& Pelletier(2011)]{2011MNRAS.417.1148L}
 Lemoine, M., \& Pelletier, G.\ 2011,
  MNRAS, 417, 1148 
%
\bibitem[Li et~al.(2003)]{2003APJ...599..380}
  Li Z., Dai Z. G., Lu T., Song L. M.,
    2003, ApJ, 599, 380
%
\bibitem[Lithwick \& Sari(2001)]{2001APJ...555..540}
  Litwick Y., Sari R.,
    2001, ApJ, 555, 540
%
\bibitem[Lytikov \& Blandford(2003)]{2003MNRAS...346..540}
  Lytikov M., Blandford R.,
    2003, MNRAS, 346, 540
%
%
\bibitem[McBreen et~al.(1994)]{1994MNRAS...271..662}
  McBreen B., Hurley K. J., Long R., Metcalfe L.,
    1994, MNRAS, 271, 662
%
\bibitem[Meegan et~al.(2009)]{2009ApJ...702..791}
  Meegan C. A., et~al.,
    2009, ApJ, 702, 791
%
\bibitem[M{\'e}sz{\'a}ros(2002)]{2002ARAA...40..137}
  M\'{e}sz{\'a}ros P.,
    2002, ARA\&A, 40, 137.
%
\bibitem[M{\'e}sz{\'a}ros(2006)]{2006RPPh...69..2259}
  M{\'e}sz{\'a}ros P.,
    2006, Reports on Prog. in Phys., 69, 2259
%
\bibitem[M{\'e}sz{\'a}ros \& Rees(1993)]{1993APJ...405..278}
  M{\'e}sz{\'a}ros P., Rees M.~J.,
    1993, ApJ, 405, 278
%
\bibitem[M{\'e}sz{\'a}ros \& Rees(2000)]{2000ApJ...530..292}
  M{\'e}sz{\'a}ros P., Rees M.~J.,
    2000, ApJ, 530, 292
%
\bibitem[M{\'e}sz{\'a}ros, Laguna \& Rees(1993)]{1993ApJ...415..181}
  M{\'e}sz{\'a}ros P., Laguna P., Rees M. J.,
    1993, ApJ, 414, 181
%
\bibitem[Metzger et~al.(1997)]{1997Nat...387..878}
  Metzger M. R., Djorgovski S. G., Kulkarni S. R., Steidel C. C., Adelberger K. L., Frail D. A., Costa E., Frontera F.,
    1997, Nature, 387, 878
%
\bibitem[Mitra(1998)]{1998ApJ...492..677}
  Mitra A.,
    1998, ApJ, 492, 677
%
\bibitem[Nakar(2007)]{2007PR...442..166}
  Nakar E.,
    2007, Physics Reports, 442, 166
%
\bibitem[Nakar \& Piran(2002)]{2002ApJ...572..L139}
  Nakar E., Piran T.,
    2002, ApJ, 572, L139
%
\bibitem[Nakar et al.(2009)]{2009ApJ...703..675N}
 Nakar, E., Ando, S., \& Sari, R.\ 2009, ApJ, 703, 675
%
\bibitem[Narayan, Paczynski \& Piran(1992)]{1992ApJ...395..L83}
  Narayan R., Paczynski B., Piran T.,
    1992, ApJ, 395, L83
%
\bibitem[Nava et al.(2011)]{2011MNRAS.415.3153N}
 Nava, L., Ghirlanda, G., Ghisellini, G., \& Celotti, A.,
 2011, MNRAS, 415, 3153
%
\bibitem[Nolan et~al.(1992)]{1992ITNS...39..993}
  Nolan P. L., et~al.,
    1992, IEEE Trans. on Nucl. Sci., 39, 993
%
%
\bibitem[Norris(2002)]{2002ApJ...579..386}
  Norris J. P.,
    2002, ApJ, 579, 386
%
\bibitem[Norris \& Bonnell(2006)]{2006ApJ...643..266}
  Norris J. P., Bonnell J. T.,
    2006, ApJ, 643, 266
%
\bibitem[Norris et~al.(1996)]{1996ApJ...459..393}
  Norris J. P., Nemiroff R. J., Bonnell J. T., Scargle J. D., Kouveliotou C., Paciesas W. S., 
  Meegan, C. A., Fishman, G. J.,
    1996, ApJ, 459, 393.
%
\bibitem[Norris, Marani \& Bonnell(2000)]{2000ApJ...534..248}
  Norris J. P., Marani G. F., Bonnell J. T.,
    2000, ApJ, 534, 248
%
\bibitem[Paciesas et~al.(2001)]{2001grba.conf...13}
  Paciesas W.~S., Preece R.~D., Briggs M.~S. Mallozzi R.~S., 2001,
  ESO Astrophysics Symposia, eds E. Costa, F. Frontera, J. Hjorth.
  Springer-Verlag, 2001, 13
%
\bibitem[Paciesas et~al.(2003)]{2003AIPC...662..248}
  Paciesas W.~S., Briggs M.~S., Preece R.~D., Mallozzi R.~S.,
    2003, in Gamma-Ray Burst and Afterglow Astronomy, eds
  Ricker G.R., Vanderspek R.K., AIP Conf. Proc., 662, 248
%
\bibitem[Paciesas et~al.(2011)]{2011ApJ}
  Paciesas W.~S., et~al.,
    2011, ApJS, in press
%
\bibitem[Paczynski \& Rhoads(1993)]{1993ApJ...418..L5}
  Paczynski B., Rhoads J.,
    1993, ApJ, 418, L5
%
\bibitem[Paczynski \& Xu(1994)]{1994ApJ...427..708}
  Paczynski B., Xu G.,
    1994, ApJ, 427, 708
%
\bibitem[Pe'er et al.(2006)]{2006ApJ...642..995P}
 Pe'er, A., M{\'e}sz{\'a}ros, P., \& Rees, M.~J.,
  2006, ApJ, 642, 995
%
\bibitem[Pe'er \& Waxman(2005)]{2005ApJ...633..1018}
  Pe'er A., Waxman E.,
   2005, ApJ, 633, 1018
%
\bibitem[Pe'er \& Ryde(2011)]{2011ApJ...732...49P}
 Pe'er, A., \& Ryde, F.,
  2011, ApJ, 732, 49
%
%
\bibitem[Peng et~al.(2011)]{2011AN...332..92}
  Peng Z. Y., Yin Y., Bi X. W., Bao Y. Y., Ma L.,
    2011, AN, 332, 92
%
\bibitem[Piran(1999)]{1999PR...314..575}
  Piran T.,
    1999, Phys. Rep., 314, 575
%
\bibitem[Piran(2003)]{2003Nat...422..268}
  Piran T.,
    2003, Nature, 422, 268
%
%
\bibitem[Piro et al.(2001)]{2001SBH}
  Piro L., et al.
   2001, in Gamma-ray bursts in the afterglow era, eds. Costa E., Frontera F., Hjorth J.,
    Springer, Berlin, Heidelberg, p. 415
%
\bibitem[Preece et~al.(1998)]{1998ApJ...506L..23P}
  Preece R.~D., Briggs M.~S., Mallozzi R.~S.,  Pendleton,G. N., Paciesas W. S., Band, D. L.,
    1998, ApJ, 506, L23
%
\bibitem[Preece et~al.(2000)]{2000APJS...126..19}
  Preece R.~D., Briggs M. S., Mallozzi R. S., Pendleton G. N., Paciesas W. S., Band, D. L., 
    2000, ApJS, 126, 19
%
\bibitem[Presani(2011)]{2011AIPC...1358..361}
  Presani E.,
    2011, in Gamma Ray Bursts 2010, eds McEnery J.E.,
   Racusin J.L., Gehrels N., AIP Conf. Proc., 1358, 361
%
\bibitem[Razzaque, M{\'e}sz{\'a}ros \&  Zhang(2005)]{2005IJMPA...20..3163}
  Razzaque S., M{\'e}sz{\'a}ros P., Zhang B.,
    2005, J. of Mod. Phys. A, 20, 3163
%
\bibitem[Razzaque, Mena \& Dermer(2009)]{2009ApJ...691L..37R}
  Razzaque S., Mena O., Dermer C.~D.,
    2009, ApJ, 691, L37
%
\bibitem[Razzaque, Dermer \& Finke(2010)] {2010OAJ.....3..150R}
  Razzaque S., Dermer C. D., Finke J. D.,
    2010, Open Astronomy Journal, 3, 150
%
\bibitem[Rees \& M{\'e}sz{\'a}ros(1992)]{1992MNRAS...258..41}
  Rees M.~J., M{\'e}sz{\'a}ros P.,
    1992, MNRAS, 258, 41
%
\bibitem[Rees \& M{\'e}sz{\'a}ros(1994)]{1994ApJ...430..L93}
  Rees M.~J., M{\'e}sz{\'a}ros P.,
    1994, ApJ, 430, L93
%
\bibitem[Rees \& M{\'e}sz{\'a}ros(2005)]{2005ApJ...628..847}
  Rees M.~J., M{\'e}sz{\'a}ros P.,
    2005, ApJ, 628, 847
%
\bibitem[Rhoads(1999)]{1999ApJ...525..737}
  Rhoads J.,
    1999, ApJ 525, 737
%
\bibitem[Ricker et~al.(2003)]{2003AIP...662..3}
  Ricker G. R., et~al.,
    2003, in Gamma-Ray Burst and Afterglow Astronomy, eds
  Ricker G.R., Vanderspek R.K., AIP Conf. Proc., 662, 3
%
\bibitem[Richardson et~al.(1996)]{1996AIPC...384..87}
  Richardson G., Koshut T., Paciesas W. S., Kouveliotou C., 
    1996, in Gamma-ray bursts: 3rd Huntsville symposium, eds, 
    Kouveliotou C., Fishman G.J., BriggsM.F.,  AIP Conf. Proc., 384, 87
%
\bibitem[Ryan(1989)]{1989NuPhS...10..121}
  Ryan J. M.,
    1989, Nucl. Phys. B Proc. Suppl., 10, 121
%
\bibitem[Ryde(2004)]{2004ApJ...614..827}
  Ryde F.,
    2004, ApJ, 614, 827
%
\bibitem[Ryde(2005a)]{2005ApJ...625L..95}
  Ryde F.,
    2005a, ApJ, 625, L95
%
\bibitem[Ryde(2005b)]{2005AAP...429..869}
  Ryde F.,
    2005b, A\&A, 429, 869
%
\bibitem[Ryde et~al.(2010)]{2010ApJ...709..172}
  Ryde F., et~al.,
    2010, ApJ, 709, L172
%
\bibitem[Ryde et~al.(2011)]{2011MNRAS...415..3693}
  Ryde F., et~al.,
    2011, MNRAS, 415, 3693
%
\bibitem[Sakamoto et al.(2006)]{2006ApJ...636L..73S}
  Sakamoto T., et al.,
    2006, ApJ, 636, L73
%
\bibitem[Salmonson(2000)]{2000ApJ...544..L115}
  Salmonson J. D.,
    2000, ApJ, 544, L115
%
\bibitem[Salmonson \& Galama(2002)]{2002ApJ...569..682}
  Salmonson J. D., Galama T. J.,
    2002, ApJ, 569, 682
%
\bibitem[Sari \& Piran(1997)]{1997ApJ...485..270}
  Sari R., Piran T.,
    1997, ApJ, 485, 270
%
\bibitem[Sari \& Piran(1999)]{1999ApJ...517..L109}
  Sari R., Piran T.,
    1999, ApJ, 517, L109
%
\bibitem[Sari \& M{\'e}sz{\'a}ros(2000)]{2000ApJ...535..L33}
  Sari R., M{\'e}sz{\'a}ros, P.,
    2000, ApJ, 535, L33
%
%
\bibitem[Schaefer(2004)]{2004ApJ...602..306}
  Schaefer B. E.,
    2004, ApJ, 602, 306
%
\bibitem[Schaefer(2007)]{2007ApJ...660..16}
  Schaefer B. E.,
    2007, ApJ, 660, 16
%
\bibitem[Sommer et~al.(1994)]{1994ApJ...422..63}
  Sommer M., et~al.,
    1994, ApJ, 422, L63
%
\bibitem[Stecker, Malkan \& Scully(2006)]{2006ApJ...648..774S}
  Stecker F.~W., Malkan M.~A., Scully S.~T.,
    2006, ApJ, 648, 774
%
\bibitem[Stecker, Malkan \& Scully(2007)]{2007ApJ...658.1392S}
  Stecker F.~W., Malkan M.~A., Scully S.~T.,
    2007, ApJ, 658, 1392
%
\bibitem[Tanvir et~al.(2009)]{2009Nat...461.1254}
  Tanvir N. R., et~al.,
    2009, Nature, 461, 1254
%
\bibitem[Thompson(2006)]{2006ApJ...651..333}
  Thompson C.,
    2006, ApJ, 651, 333
%
\bibitem[Toma et al.(2011)]{2011MNRAS.415.1663T}
Toma, K., Wu, X.-F., \& M{\'e}sz{\'a}ros, P.,
 2011, MNRAS, 415, 1663
%
\bibitem[Ukwatta et~al.(2011)]{2011MNRAS...410}
  Ukwatta T. N., et~al.,
    2011, MNRAS, in press
%
\bibitem[Urata et~al.(2007)]{2007ApJ...668L..95}
  Urata Y., et~al.,
    2007, ApJ, 668, L95
%
\bibitem[Vanderspek et~al.(2004)]{2004AIP...727..57}
  Vanderspek R., et~al.,
    2004, in Gamma-ray burst symposium, eds, 
    Fenimore E., Galassi M.,  AIP Conf. Proc., 727, 57
%
\bibitem[van Eerten, Zhang \& MacFadyen(2010)]{2010ApJ...722..235}
  van Eerten H. J., Zhang W., MacFadyen A.,
    2010, ApJ, 722, 235
%
\bibitem[van Eerten et al.(2011)]{2011MNRAS...410..2016}
  van Eerten H. J., Meliani Z., Wijers R.~A.~M.~J., Keppens R.,
    2011, MNRAS, 410, 2016
%
\bibitem[van Paradijs et~al.(1997)]{1997Nat...386..686}
  van Paradijs J., et~al.,
    1997, Nature, 386, 686
%
\bibitem[van Paradijs, Kouveliotou \& Wijers(2000)]{2000ARAA...38..379}
  van Paradijs J., Kouveliotou C., Wijers, R.~A.~M.~J.,
    2000, ARA\&A, 38, 379
%
\bibitem[Vieregg et~al.(2011)]{2011ApJ...736...50}
  Vieregg A. G. et~al.,
    2011, ApJ, 736, 50
%
\bibitem[Vietri, De Marco \& Guetta(2003)]{2003ApJ...592..378}
  Vietri M., De Marco D., Guetta D.,
    2003, ApJ, 736, 50
%
\bibitem[Virgili et~al.(2009)]{2009MNRAS.392...91}
  Virgili F. J., Zhang B., Troja E., O'Brien P.,
    2009, MNRAS, 392, 91
%
\bibitem[Waxman(1995)]{1995PRL...75..386}
  Waxman E.,
    1995, Phys. Rev. Lett., 75, 386
%
\bibitem[Waxman(2003)]{2003LNP...598..393}
  Waxman E.,
    2003, Lect. Notes in Phys. 598, 393
%
\bibitem[Waxman \& Bachall(1997)]{1997PRL...78..2292}
  Waxman E., Bachall J.,
    1997, Phys. Rev. Lett., 78, 2292
%
\bibitem[Zeh, Klose \& Kann(2006)]{2006ApJ...637..889}
  Zeh A., Klose S., Kann D.~A., 2006, ApJ, 637, 889 
%
\bibitem[Zhang et~al.(2006)]{2006MNRAS...373..729}
  Zhang Z., Xie G. Z., Deng J. G., Jin W.,
    2006, MNRAS, 373, 729
%
\bibitem[Zhang et~al.(2006)]{2006ApJ...642..354}
  Zhang B., Fan Y. Z., Dyks J., Kobayashi S., Meszaros P., Burrows D. N.,
  Nousek J. A., Gehrels N.,
    2006, ApJ, 642, 354
%
\bibitem[Zhang \& Pe'er(2009)]{2009ApJ...700L..65Z}
 Zhang, B., \& Pe'er, A.,
 2009, ApJL, 700, L65
%
\bibitem[Zhang \& Yan(2011)]{2011ApJ...726...90Z}
 Zhang, B., \& Yan, H.,
  2011, ApJ, 726, 90
%
\bibitem[Zhang et al.(2011)]{2011ApJ...730..141Z}
 Zhang, B.-B., Zhang, B., Liang, E.-W., et al., 
  2011, ApJ, 730, 141
%

\end{thebibliography}
\end{document}